\documentclass[traditabstract]{aa}
\usepackage{graphicx}
\usepackage[varg]{txfonts}
\usepackage{hyperref}
\usepackage{longtable,lscape}
\usepackage{natbib}
\bibpunct{(}{)}{;}{a}{}{,}

\def\ud{\,\mathrm{d}}

\begin{document}
\title{Isolated ellipticals and their globular cluster systems}
\subtitle{III. NGC 2271, NGC 2865, NGC 3962, NGC 4240 and IC 4889
  \thanks{Based on observations obtained at the Gemini Observatory,
    which is operated by the Association of Universities for Research
    in Astronomy, Inc., under a cooperative agreement with the NSF on
    behalf of the Gemini partnership: the National Science Foundation
    (United States), the Science and Technology Facilities Council
    (United Kingdom), the National Research Council (Canada), CONICYT
    (Chile), the Australian Research Council (Australia), Minist\'erio
    da Ci\^encia, Tecnologia e Inova\c{c}\~ao (Brazil) and Ministerio
    de Ciencia, Tecnolog\'ia e Innovaci\'on Productiva
    (Argentina).}}

\author{
R. Salinas \inst{\ref{inst:finca},\ref{inst:msu}}
\and
A.Alabi\inst{\ref{inst:tuorla},\ref{inst:swinburne}} 
\and
T. Richtler \inst{\ref{inst:udec}}
\and
R. R. Lane \inst{\ref{inst:udec}}  
}

\institute{ Finnish Centre for Astronomy with ESO (FINCA), University
  of Turku, V\"ais\"al\"antie 20, FI-21500 Piikki\"o,
  Finland\label{inst:finca} 
\and
Department of Physics and Astronomy, Michigan State University, East
Lansing, MI 48824, USA\label{inst:msu} 
\and
Tuorla Observatory, University of Turku, V\"ais\"al\"antie 20,
FI-21500 Piikki\"o, Finland\label{inst:tuorla}
\and
Centre for Astrophysics and Supercomputing, Swinburne University of
Technology, Hawthorn, VIC 3122, Australia\label{inst:swinburne}
\and
Departamento de Astronom\'{\i}a, Universidad de Concepci\'on,
Concepci\'on, Chile\label{inst:udec}
}

\date{Accepted 21 Feb 2015}

\abstract{ As tracers of star formation, galaxy assembly and mass
  distribution, globular clusters have provided important clues to our
  understanding of early-type galaxies. But their study has been
  mostly constrained to galaxy groups and clusters where early-type
  galaxies dominate, leaving the properties of the globular cluster
  systems (GCSs) of isolated ellipticals as a mostly uncharted
  territory. We present Gemini-South/GMOS $g'i'$ observations of five
  isolated elliptical galaxies: NGC 3962, NGC 2865, IC 4889, NGC 2271
  and NGC 4240.  Photometry of their GCSs reveals clear color
  bimodality in three of them, remaining inconclusive for the other
  two. All the studied GCSs are rather poor with a mean specific
  frequency $S_N\sim 1.5$, independently of the parent galaxy
  luminosity. Considering also previous work, it is clear that
  bimodality and especially the presence of a significant, even
  dominant, population of blue clusters occurs at even the most
  isolated systems, casting doubts on a possible accreted origin of
  metal-poor clusters as suggested by some models. Additionally, we
  discuss the possible existence of ultra-compact dwarfs around the
  isolated elliptical NGC 3962.}

\titlerunning{Globular clusters in isolated ellipticals II}
\authorrunning{Salinas et al.}
\maketitle

\section{Introduction}
\label{sec:intro}

The globular cluster systems (GCSs) of old, massive elliptical
galaxies present an almost ubiquitous optical color bimodality
\citep[e.g.][]{kundu01,larsen01}, thought as corresponding to a
metallicity bimodality \citep[e.g.][]{strader07,usher12,brodie12},
with caveats coming from possible non-linearities of the
color-metallicity transformations introduced by horizontal-branch
stars \citep{richtler06,yoon06,chies-santos12,blakeslee12,richtler13}.

This ``universal'' bimodality has been obviously one of the main
aspects the theories of GCS formation need to address. Many scenarios,
not necessarily exclusive and somewhat overlapping, have been put
forward over the years
\citep[e.g.][]{ashman92,forbes97,cote98,beasley02,rhode05,muratov10,elmegreen12,tonini13}.

Since elliptical galaxies are known to inhabit mostly high-density
environments \citep[e.g.][]{dressler80}, most of the GCS studies, and
hence the observational constraints to these theories, have been
naturally focused on galaxy clusters
\citep[e.g.][]{peng06,harris06,strader06,liu11}.

In the current reigning paradigm of galaxy formation, accretion and
merging play the central roles \citep[e.g.][]{cole94,delucia06},
finding strong observational support
\citep[e.g.][]{ibata94,tal09,vandokkum10}. Simulations predict a
different accretion history for elliptical galaxies in low-density
environments compared to their high-density siblings
\citep[e.g.][]{niemi10}; but the impact that a low-density environment
may have on the properties of a GCS has seldom been investigated
\citep[e.g][]{gebhardt99}, although it could give important evidence
to discriminate between accretion driven
\citep[e.g.][]{cote98,hilker99} and \textit{in situ} \citep{forbes97}
scenarios.

Only a handful of isolated elliptical galaxies (IEs) have their GCS
thoroughly studied \citep[e.g.][]{spitler08}, given their paucity in
the local environment. No field ellipticals are listed, for example,
in the big compilation of \citet[][see their Table 1]{brodie06}, and
only a handful can be seen in the newer compilation of
\citet{harris13}. Recently, \citet{cho12} studied 10 early-type
galaxies in low density environments in the magnitude range
$-18.5<M_B<-20$ using HST/ACS, although solely using the
\citet{tully88} density parameter as isolation criterion.

\begin{table*}
  \caption{Basic data of the galaxies presented in this work.}
\label{table:sample}
\centering
\begin{tabular}{@{}llllllllll@{}}
  \hline\hline
  Name     &  RA        & Dec          & $l$         & $b$          &Type  & $B^T_0$   &$(m-M)_0$ &$M_B$      & fov  \\
  &     &     &      &    &     &(mag)  & (mag)    & (mag)    & (kpc) \\
  \hline
  NGC 2271 & 06:42:53.0 & --23:28:34.0 & 233:13:30.5 & --12:14:45.8 & E/S0    &12.10 &32.51&$-20.41$  & 50.9\\
  NGC 2865 & 09:23:30.2 & --23:09:41.0 & 252:57:12.4 & +18:56:29.9  & E3      &12.18 &32.89&$-20.71$  & 60.5\\
  NGC 3962 & 11:54:40.1 & --13:58:30.0 & 282:39:10.7 & +46:38:57.5  & E1      &11.59 &32.74\tablefootmark{a}&$-21.15$  &56.5\\
  NGC 4240 & 12:17:24.3 & --09:57:06.0 & 289:14:11.8 & +52:00:46.3  & E/cD?   &14.01 &32.07&$-18.06$  & 41.5\\
  IC 4889  & 19:45:15.1 & --54:20:39.0 & 343:32:15.1 & --29:25:10.4 & E5-6/S0 &11.91 &32.33&$-20.42$  & 46.8\\
  \hline
\end{tabular}
\tablefoot{Classification type comes from NED and HyperLeda databases,
  while total apparent $B$ magnitudes are taken from RC3, except for NGC 4240, taken from HyperLeda. Distance moduli have
  been adopted from \citet{tonry01}, except the one for NGC 4240
  which comes from \citet{reda04}. The last column gives the
  physical size covered by the GMOS FOV at each galaxy's distance.}
\tablefoottext{a}{Considered as uncertain by \citet{tonry01}.}
\end{table*}

This paper is part of an effort to understand the properties of GCS in
IEs and their connections to their parent galaxies, collecting a
large sample. This paper is a continuation of \citet[][hereafter Paper
  I]{lane13} who studied the GCS of the IE NGC 3585 and NGC 5812
using Washington photometry, and \citet[][Paper II]{richtler15} who
studied the IE NGC 7796. It is also complemented by the study of the
field elliptical NGC 7507 \citep{caso13} also conducted by our group.

\subsection{The galaxy sample}
\label{sec:sample}

As isolated galaxies give the opportunity to study galaxies without
the interference of many of the processes that affect their structure
and evolution in more crowded environments, many samples of IEs have
been constructed, using diverse isolation criteria
\citep{kuntschner02,
  smith04,stocke04,reda04,collobert06,nigoche07,fuse12}.

In this work we present observations of five IEs taken from the
catalogues of \citet{smith04} and \citet{reda04}; closer than $\sim$
50 Mpc and observable from the southern hemisphere.  The GCS of NGC
2865 and IC 4889 have been studied before, albeit with a smaller field
of view (see below). In the following we give their main
characteristics as found in the literature.

\begin{itemize}

\item{\bf NGC 2271} is a E/S0 at a surface brightness fluctuation
  distance of 31.8 Mpc \citep{tonry01} present in the catalogue of
  \citet{reda04}. \citet{reda07} measured an age of $11.5\pm0.5$ Gyr,
  with no evidence of radial age or metallicity gradients, while
  \citet{hau06} found strong rotation and kinematic signatures of an
  inner disk, which is also found on optical imaging
  \citep{desouza04}.

\item{\bf NGC 2865} is a E3 listed in the \citet{reda04} catalogue. It
  is known for the wealth of substructure (shells, tails, etc.) it
  possesses \citep{michard04,desouza04,tal09,urrutia14}, being most
  probably a merger remnant \citep{hau99}, with a kinematically
  distinct core \citep{hau06}. It presents a remarkably steep age
  gradient from a central $\sim1.5$ Gyr \citep{sanchez07,reda07} to
  $\sim$ 10 Gyr close to 1$R_e$; exhibiting also a steep metallicity
  gradient \citep{reda07}. Its GCS was studied by \citet{sikkema06}
  using HST/ACS, finding a sizeable blue population consistent with a
  young/intermediate GC population \citep[see also][]{trancho14}.

  \item{\bf NGC 3962} is part of the \citet{smith04} sample of field
  ellipticals, with no companions as confirmed by \citet{madore04},
  and no signs of any tidal disturbance
  \citep{tal09}. \citet{annibali07} measured an age of 10 Gyr, while
  \citet{serra10} give an age of $\sim$2.5 Gyr.
  
 \item{\bf NGC 4240} was taken from the \citet{reda04}
  catalogue. \citet{reda07} found a central age of 7.5 Gyr followed by
  a mild age gradient. The existence of an also shallow metallicity
  gradient is explained as indication of a past major merger. It
  possesses a kinematically distinct core and a mostly flat velocity
  dispersion profile \citep{hau06}.
 
 \item{\bf IC 4889} is an E/S0 \citep{laurikainen11} present in the
  \citet{smith04} catalogue, showing isophotal twisting \citep{tal09}
  as well as a dusty disk \citep{smith04}. \citet{serra10} estimate an
  age of $\sim$ 3.6 Gyr. Its GCS has been studied by
  \citet{gebhardt99} using HST data, without finding evidence for
  bimodality.

\end{itemize}

The rest of the paper is distributed as follows: in
Sect. \ref{sec:obs} we describe the observations, as well as the data
reduction and stellar photometry. In Sect. \ref{sec:color}, we present
the results on the color distribution of globular clusters, while
radial profiles are presented in Sect. \ref{sec:radial}. GC luminosity
functions are calculated in
Sect. \ref{sec:lumfun}. Sect. \ref{sec:galaxies} describes properties
of the galaxies themselves and their connections to their
GCS. Sect. \ref{sec:discussion} put the results into a broader
context, before giving the conclusions in Sect. \ref{sec:conclusions}.

\section{Observations and data reduction}
\subsection{Gemini-S/GMOS imaging}
\label{sec:obs}

\begin{table}
  \caption{Observations log}
  \label{table:obslog}
  \centering
  \begin{tabular}{@{}llcccc@{}}
  \hline\hline
Name   &  Obs. date & \multicolumn{2}{c}{Exp. time (s)} & \multicolumn{2}{c}{FWH
M (\arcsec)} \\
            &  (dd.mm.yyyy)  &   $g'$     & $i'$ & $g'$ & $i'$   \\
\hline
\noalign{\smallskip}
 NGC 2271 & 01.02.2012& $9\times 419$&$9\times 315$& 0.73 & 0.74\\
 NGC 2865 & 25.01.2012& $6\times 480$&$6\times 340$& 0.63 & 0.50\\
          & 26.01.2012& $6\times 490$&$6\times 350$& \\ 
 NGC 3962 & 22.02.2012& $6\times 610$&$6\times 440$& 0.73 & 0.60\\
 NGC 4240 & 02.02.2012& $5\times 422$&$5\times 275$& 0.80 & 0.58\\
 IC 4889  & 17.04.2012& $3\times 575$&$5\times 355$& --   & 0.75\\
          & 01.05.2012& $3\times 575$&$1\times 355$& 0.49 & --\\
 \hline
\end{tabular}
\tablefoot{The FWHM was measured on the combined images.}
\end{table}

Imaging of the five galaxies was obtained using GMOS, mounted at the
Gemini South Telescope, Cerro Pach\'on, Chile (program ID
GS-2012A-Q-6). GMOS comprises three detectors covering in total a
field-of-view of $5.5\times 5.5$ arcmin$^2$, with a pixel scale of
0.146\arcsec pixel$^{-1}$ after a $2\times 2$ binning. A set of
dithered images in the SDSS filters $g'$ and $i'$ was
obtained. Details of the observations can be seen in Table
\ref{table:obslog}.

\begin{table*}
  \caption{Transformation equation coefficients for Eqs. \ref{eq:1}
    and \ref{eq:2} for the different galaxies.}
\label{table:coeff}
 \centering
  \begin{tabular}{@{}lccccccc@{}}
  \hline\hline
Galaxy   &   \multicolumn{2}{c}{Zero-point} & \multicolumn{2}{c}{Color term} &N$_{\rm stars}$ & \multicolumn{2}{c}{rms} \\
 & $g'$ & $i'$ & $g'$ & $i'$& &$g'$&$i'$\\
\hline
 NGC 2271 & 28.226 $\pm$ 0.005& 27.900 $\pm$ 0.007 & $-0.052 \pm 
0.010$ &$-0.024 \pm 0.008$ & 11 & 0.035 & 0.028\\
  NGC 2865 & 28.253 $\pm$ 0.022& 27.917 $\pm$ 0.012 & $-0.052 \pm 
0.010$ &$-0.024 \pm 0.008$ & 15 & 0.046 & 0.041\\
NGC 3962 & 28.110 $\pm$ 0.034& 27.770 $\pm$ 0.033 & $-0.052 \pm 
0.010$ &$-0.024 \pm 0.008$ & 10 & 0.101 & 0.103\\
 NGC 4240 & 28.240 $\pm$ 0.008& 27.930 $\pm$ 0.007 & $-0.052 \pm 
0.010$ &$-0.024 \pm 0.008$ & 33 & 0.023 & 0.021\\
IC 4889  & 28.209 
$\pm$ 0.014& 27.923 $\pm$ 0.009 & $-0.007 \pm 0.014$& $ 0.001 \pm 
0.009$ & 20 & 0.031 & 0.017\\
\hline
\end{tabular}
\tablefoot{Extinction coefficients, taken from the Observatory webpages, are fixed for
  all equations with values $K_{g'}=0.18$ and $K_{i'}=0.08$. Color
  terms for NGC 2271, NGC 2865 and NGC 3962, were not fitted, but
  fixed; see main text for details. Additionally, the last four
  columns indicate parameters for Eq. \ref{eq:interp} for both
  filters as described in Sect. \ref{sec:complete}.}
\end{table*}

Image reduction was carried out inside the
Gemini/\textsc{iraf}\footnote{IRAF is distributed by the National
  Optical Astronomy Observatory, which is operated by the Association
  of Universities for Research in Astronomy (AURA) under cooperative
  agreement with the National Science Foundation.} package (v
1.9). Reduction included bias subtraction, flat-fielding, detector
mosaicing, fringe removal (for the $i'$ exposures) and image
co-addition. Processed bias, flat and fringe frames were retrieved
from the Gemini Science Archive website. The FWHM measured on point
sources in the final images can be seen in Table \ref{table:obslog}.
Image quality in the IC 4889 exposures showed considerable variations
between the two dates, for this reason the combined $g'$ image takes
data only from the May 2012 observations, while the combined $i'$ was
constructed with images coming from the night in April 2012 only. 

\subsubsection{Stellar photometry}
\label{sec:photometry}

In order to ease the source detection, images were first median
filtered using a box of 10\arcsec\, aside. This process removes the
parent galaxy light (everywhere but in the innermost arcseconds),
leaving the point sources unaltered. A median value of the sky,
determined from blank portions in the unsubtracted images distant from
the central galaxy, was then added back to the images to keep
the right photon statistics during the detection and photometry
stages.

Object identification was carried out using SE\textsc{xtractor}
\citep{bertin96}. The choice of the detection filter can have
important consequences on the number of identified sources; while a
Gaussian filter provides a better detection rate for the faintest
sources, a ``mexican hat'' filter gives better results for sources
very close to saturated stars or to the center of the galaxy
unsubtracted residual. Given the relatively low Galactic latitude of
some of the targets (NGC 2271 and NGC 2865) and the clumpy nature of
some galaxy residuals (for example, IC 4889, see
Sect. \ref{sec:galaxies}), the ``mexican hat'' filter is the most
suitable option. SE\textsc{xtractor} was then run twice on each image
using both detection filters. Duplicate detections within 3 pixels
were removed from the merged catalogues. 

Concentric aperture and psf-fitting photometry was obtained using the
stand-alone version of \textsc{daophot/allstar} \citep{stetson87}
using as input the object coordinates from
SE\textsc{xtractor}. Between 30 and 50 bright isolated stars per field
were chosen to determine the psf shape. Aperture corrections for each
frame were measured using a subset of these stars.

Point sources were selected based on the behaviour of the
\textsc{SExtractor} output parameter \verb+class_star+ and
\textsc{allstar} parameter \verb+sharp+. The accepted range of values
for both parameters were determined from the results of the artificial
stars experiments (see Sect.  \ref{sec:complete}). Final cross-search
between the stellar positions in the two different bands was done
using \textsc{daomatch/daomaster} \citep{stetson93}, keeping into the
final catalogue only sources present in both bands with a tolerance of
3 pixels ($\sim 0.5\arcsec$) in their positions, thus weeding out
remaining spurious detections.

Standard star frames from the southern extension of the
\citet{smith02} catalogue observed on the same nights than our science
frames were retrieved from the Gemini Science Archive. These frames
were reduced following the same procedure of the science images, with
the exception of the fringe correction, which was deemed to be
unnecessary given the short exposures ($\sim$ 5 sec) in the $i'$
filter. The number of standard stars measured for each night can be
seen in Table \ref{table:coeff}. The transformation equations to
the standard system used were
\begin{equation}\label{eq:1}
g'_{\rm obs} = g'_{\rm std} + Z_g + K_g (X-1) + b_g (g'-i')_{\rm std} 
\end{equation}
\begin{equation}\label{eq:2}
i'_{\rm obs} = i'_{\rm std} + Z_i + K_i (X-1) + b_i (g'-i')_{\rm std} 
\end{equation}
where $X$ denotes the airmass of the observation. Coefficients were
obtained using \textsc{iraf/photcal}. The extinction coefficients were
taken from the Gemini webpages and not fit, given the small range of
airmasses covered by the observations. Also, when less than 20 stars
were available on a field, the color terms were fixed to the values
obtained for NGC 4240 for which the largest number of standard stars
was available. Coefficients used for each galaxy can be seen in Table
\ref{table:coeff}.

\subsubsection{Photometry completeness}
\label{sec:complete}

\begin{figure}
\includegraphics[width=0.49\textwidth]{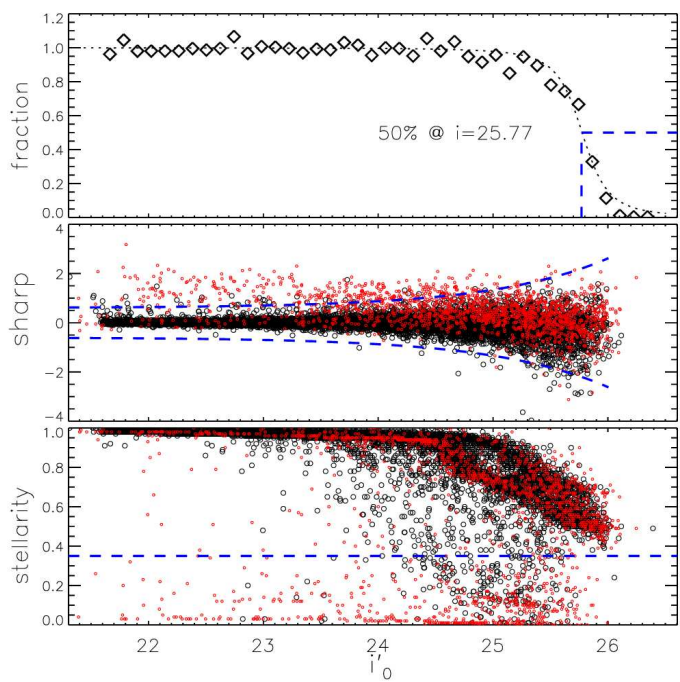}
\caption{{\it Top panel}: Fraction of recovered stars after the
  artificial stars experiment in the $i'$ observations of NGC
  4240. The dotted line is the fitted interpolation function (see
  Eq. \ref{eq:interp}). The dashed blue lines indicate the 50\% level
  of completeness. {\it Middle panel}: Behaviour of the
  \textsc{allstar} parameter sharp. {\it Bottom panel}:
  \textsc{SExtractor} stellarity index. In both panels artificial
  stars are indicated with black circles, while the real sources are
  plotted in red. The blue dashed lines indicate the point source
  criteria applied to the science data based on the artificial one.}
\label{fig:completeness}
\end{figure}

\begin{table}
  \caption{Completeness}
\label{table:complete}
 \centering
  \begin{tabular}{@{}lcccc@{}}
  \hline\hline
\multicolumn{1}{c}{Galaxy} &\multicolumn{2}{c}{$\alpha$} &\multicolumn{2}{c}{$m_{50}$}\\
&$g'$&$i'$&$g'$&$i'$\\
\hline
\noalign{\smallskip}
 NGC 2271 &1.32&1.63&26.17&25.23\\
 NGC 2865 &1.90&2.95&27.32 &26.24\\
 NGC 3962 &2.06 &3.04&26.71&25.42\\
 NGC 4240 &2.41 &4.24&26.45&25.77\\
 IC 4889  &3.29 &2.25&26.82&24.41\\
\hline
\end{tabular}
\end{table}

To study the level of completeness as a function of magnitude,
artificial stars experiments were conducted using
\textsc{daophot/addstar}. In each of the co-added frames (i.e. the
ones where the galaxy light was not subtracted yet), we added 200
point sources covering a wide magnitude range. These new frames were
then processed in the same way previously described: galaxy
subtraction with median filtering, source detection with
\textsc{SExtractor} and psf photometry with \textsc{daophot}. The
procedure was repeated 50 times for a total of 10000 artificial stars
per galaxy per filter. An example of the fraction of recovered stars
as a function of magnitude can be seen in Fig. \ref{fig:completeness}
(top panel).

The fraction of recovered stars was fitted with the interpolation formula,

\begin{equation}
\label{eq:interp}
f = \frac{1}{2}\left(1-\frac{\alpha(m-m_{50})}{\sqrt{1+\alpha^2(m-m_{50})^2}}\right)
\end{equation}
which gives a good analytical description of the completeness, and
have been used for many GC studies
\citep[e.g.][]{mclaughlin94,fleming95,wehner08,alamo12}. The free
parameters $\alpha$ and $m_{50}$ indicate the speed at which the
detection rate falls at the faint level, and the 50\% completeness,
respectively.  The values for both parameters for all fields and
filters can be seen in Table \ref{table:complete}.

Artificial stars experiments are not only useful to understand the
detection limits, but also to understand the behaviour of the
photometric measurements at these faint magnitudes. Separation between
point and extended sources is a pervading problem in galactic and
extragalactic studies and several methods are commonly used
\citep[e.g.][]{harris91,bertin96,clem08,durret09,cho12}. In our case
we preferred to use a combination of the \textsc{SExtractor} output
parameter \verb+class_star+, a stellarity likelihood, and the
\textsc{allstar} parameter \verb+sharp+, which measures deviations
from the derived psf. In Fig. \ref{fig:completeness}, we show an
example of the behaviour of both parameters as a function of magnitude
for the artificial stars (black circles) and the detected sources (red
circles). Blue dashed lines show the limits that define the envelopes
of the artificial stars. Only about 1\%--2\% of the total artificial
stars lie beyond these limits. In the following we consider as point
sources, the sources that satisfy these criteria.

\subsubsection{Estimating the contamination}
\label{sec:contamination}

GCS of elliptical galaxies can have radial extensions up to hundreds
of kpc \citep[e.g.][]{richtler11,schuberth12}. Given the relatively
small FOV provided by GMOS at the galaxies' distance (see column 10 in
Table \ref{table:sample}), a control field cannot be taken within the
same telescope pointing as in wide-field studies \citep[e.g.][Paper
I]{rhode01,dirsch03}, but has to be selected elsewhere. Control fields
are necessary in order to determine the contamination the GC samples
suffer; contamination coming partly from stars and from faint
background galaxies that cannot be culled during the point-source
selection (Sect. \ref{sec:photometry}).

Since our observing program did not include any control field, an
appropriate field was searched in the Gemini Science Archive, finding
that the William Herschel Deep Field (WHDF) had similar total 
exposure times
and image quality than our science data, and a very similar setup,
using $g'$ and $i'$ filters on Gemini-N/GMOS \citep[Gemini ID
  GN-2001B-SV-104,][]{metcalfe01}. The main caveat of using the WHDF
as control field lies on its relatively high Galactic latitude
($b\sim-61\degr$), implying a likely underestimation of the stellar
contamination, although at the faint levels where the bulk of the GC
studied here are located, the dominant component of the contamination
will be unresolved background galaxies, instead of stars
\citep{faifer11}. The WHDF counts were hence complemented by adding
star counts coming from the Besan\c{c}on Galactic models
\citep{robin03} at each galaxies' position. As noted by 
\citet{lane11}, comparisons between the Besan\c{c}on models and 
pencil beams surveys can be problematic, due to stellar streams or 
other Local Group objects that are not included in the model. In 
our case, neither the WHDF nor the studied galaxies lie close to any 
of these features.

\begin{figure*}
\centering
\begin{tabular}{cc}
\includegraphics[width=0.42\linewidth]{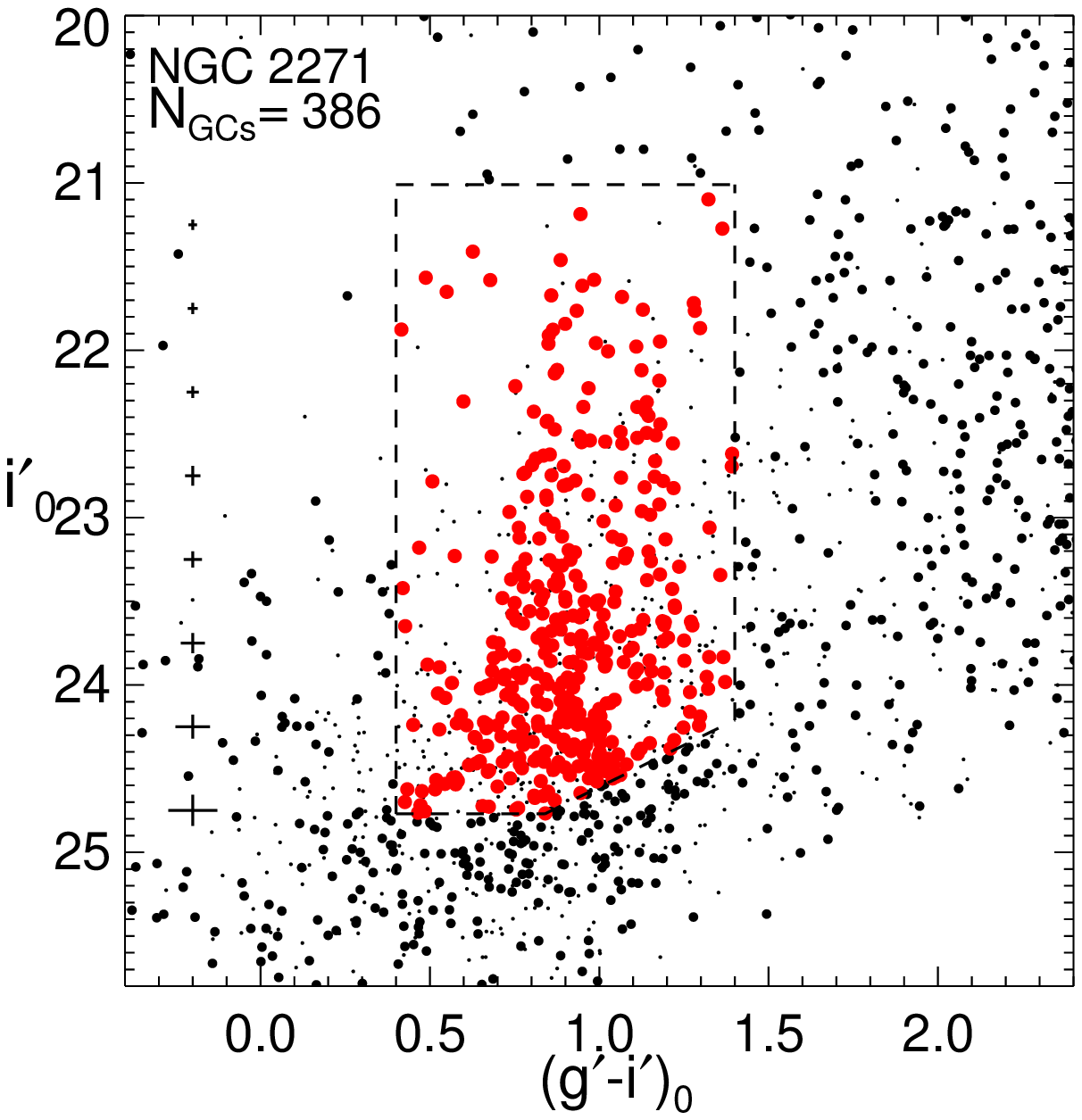}&
\includegraphics[width=0.42\linewidth]{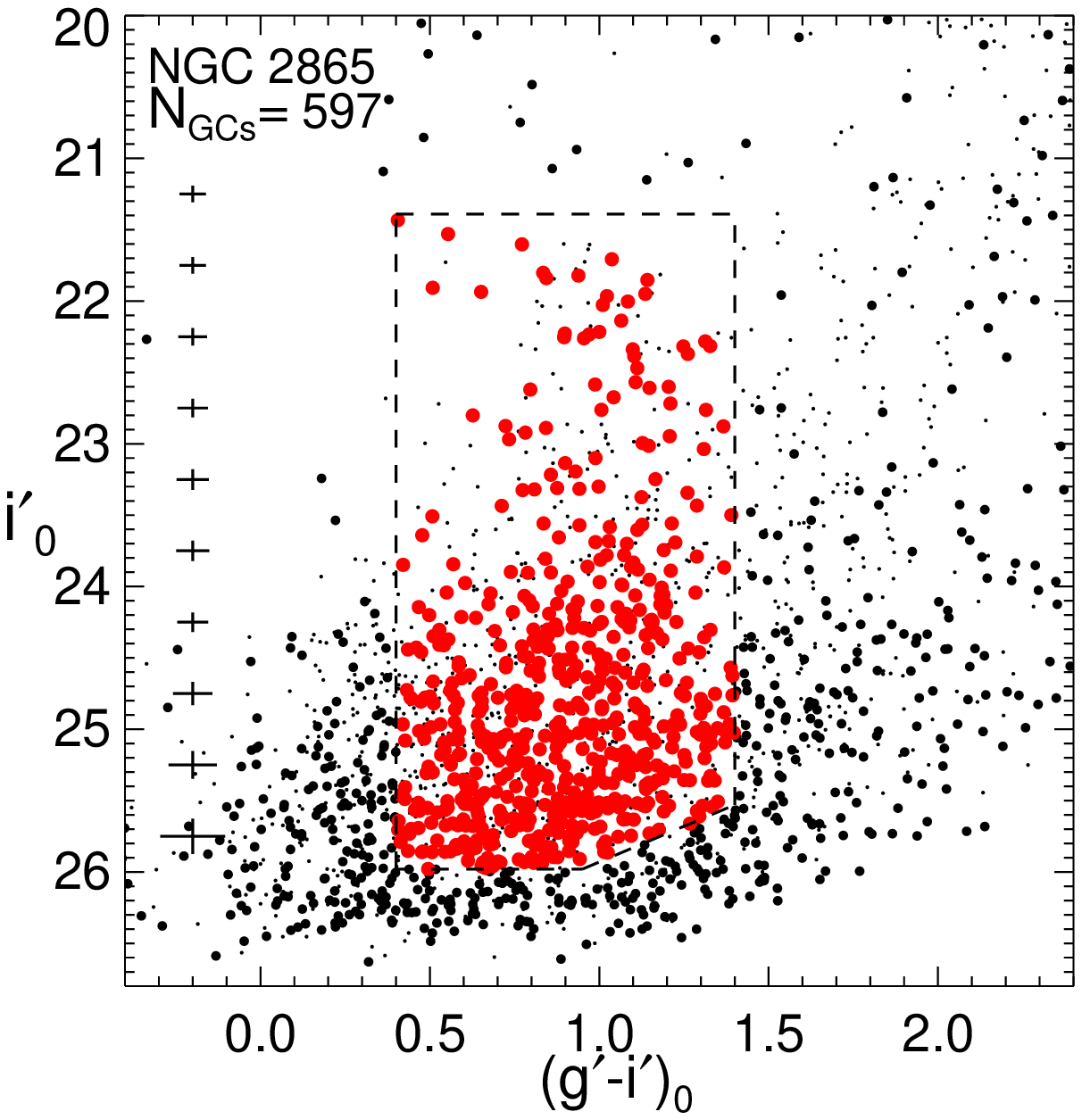}\\
\includegraphics[width=0.42\linewidth]{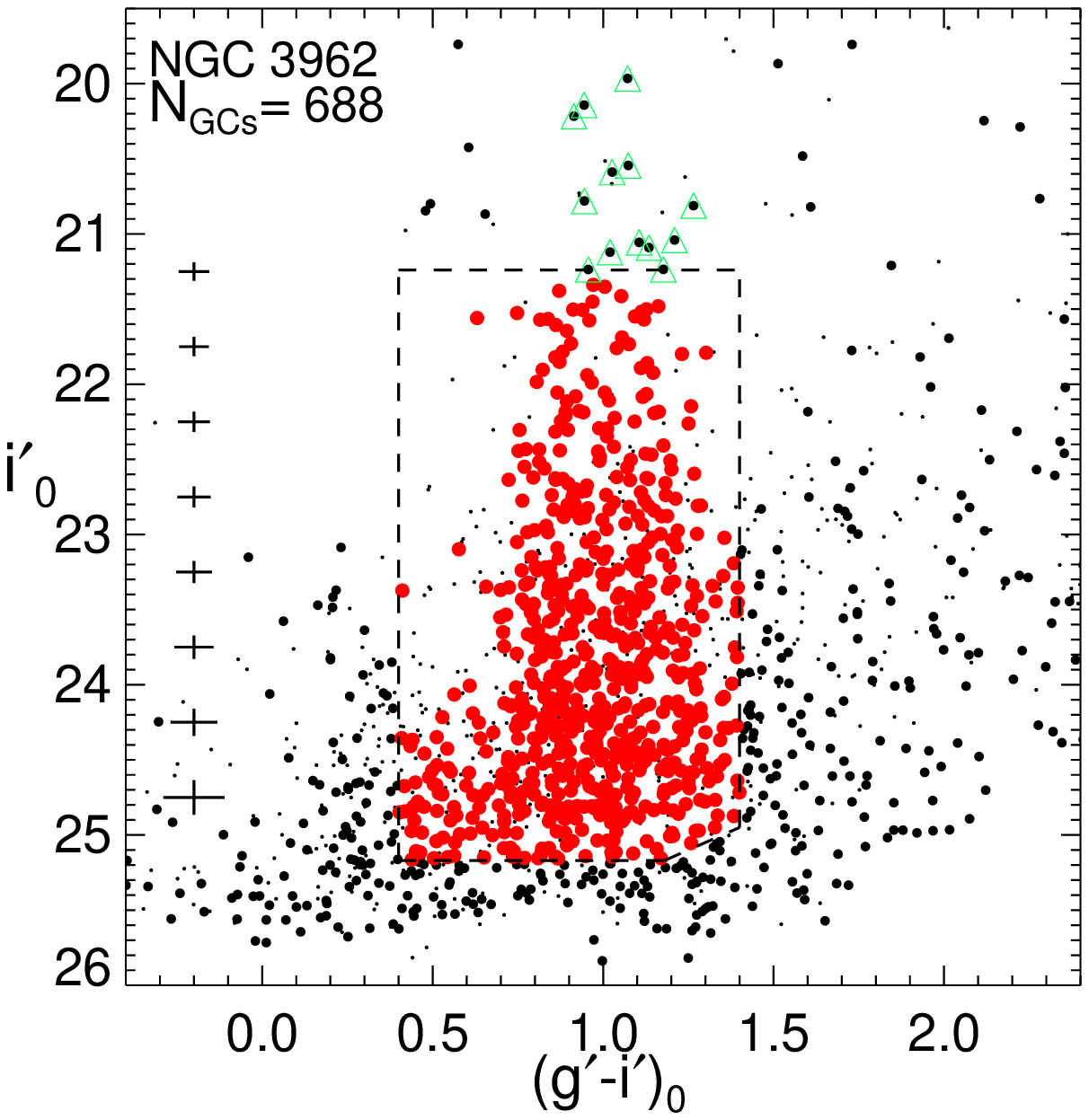}& 
\includegraphics[width=0.42\linewidth]{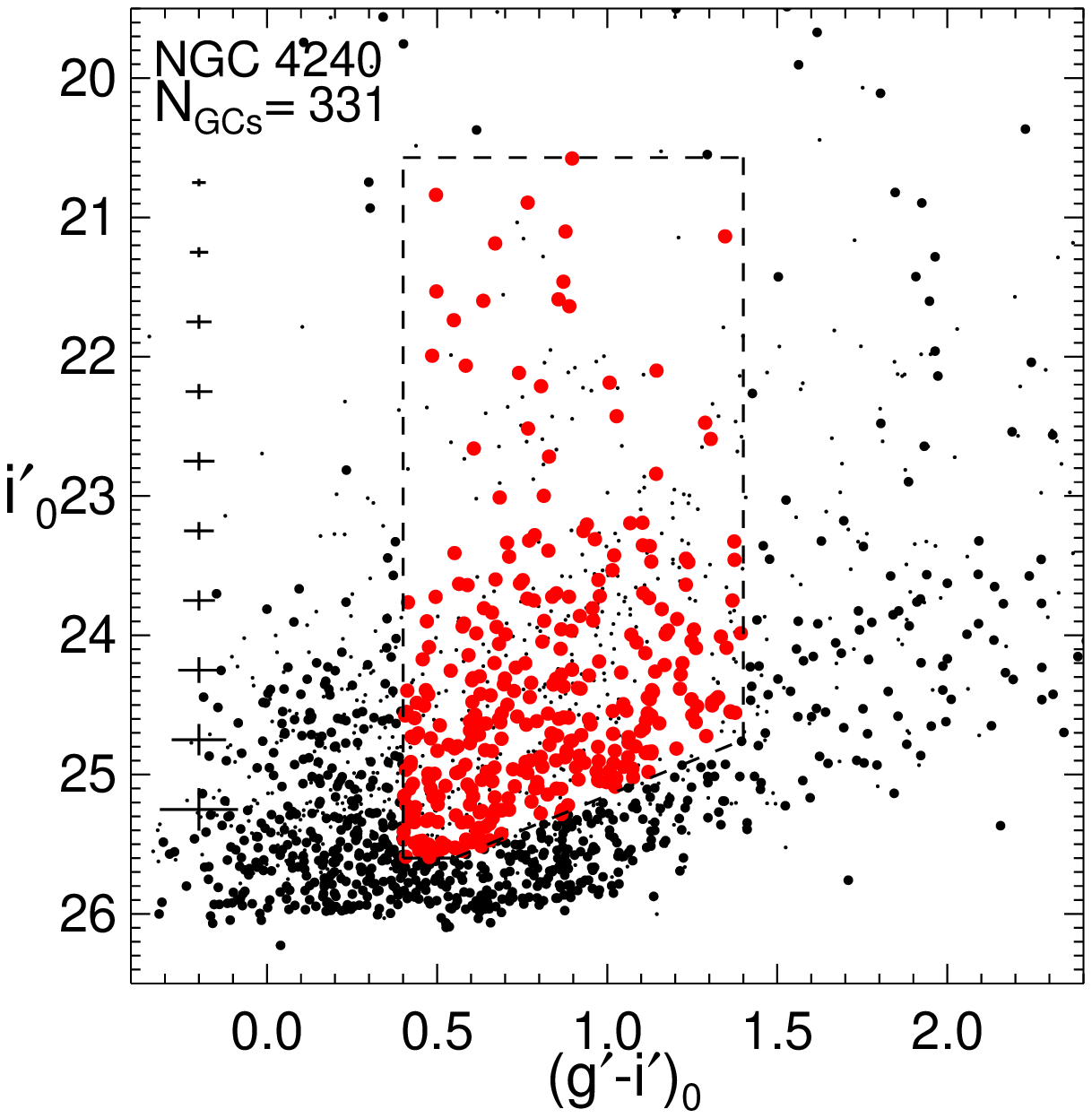}\\
\includegraphics[width=0.42\linewidth]{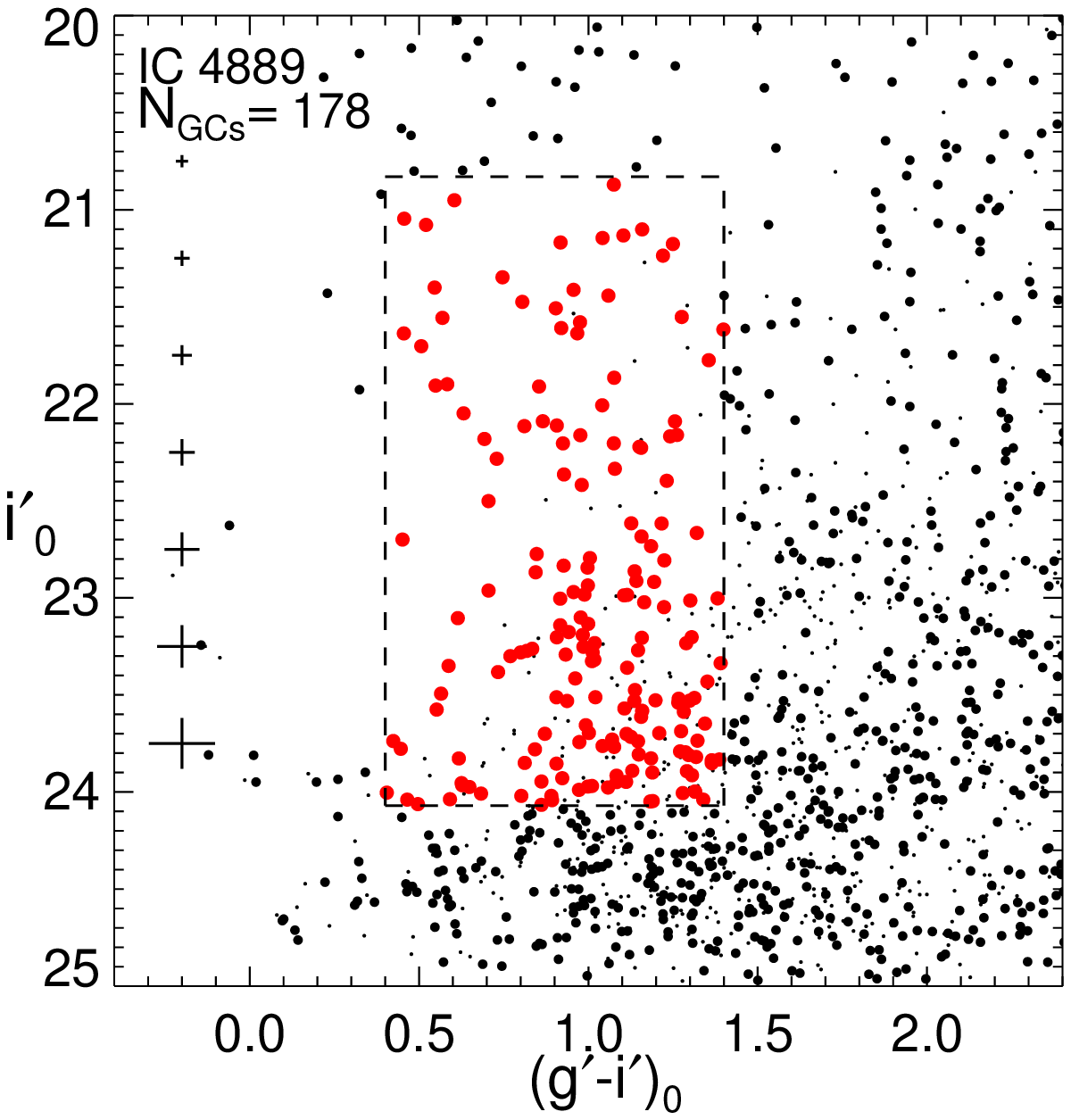} &
\includegraphics[width=0.42\linewidth]{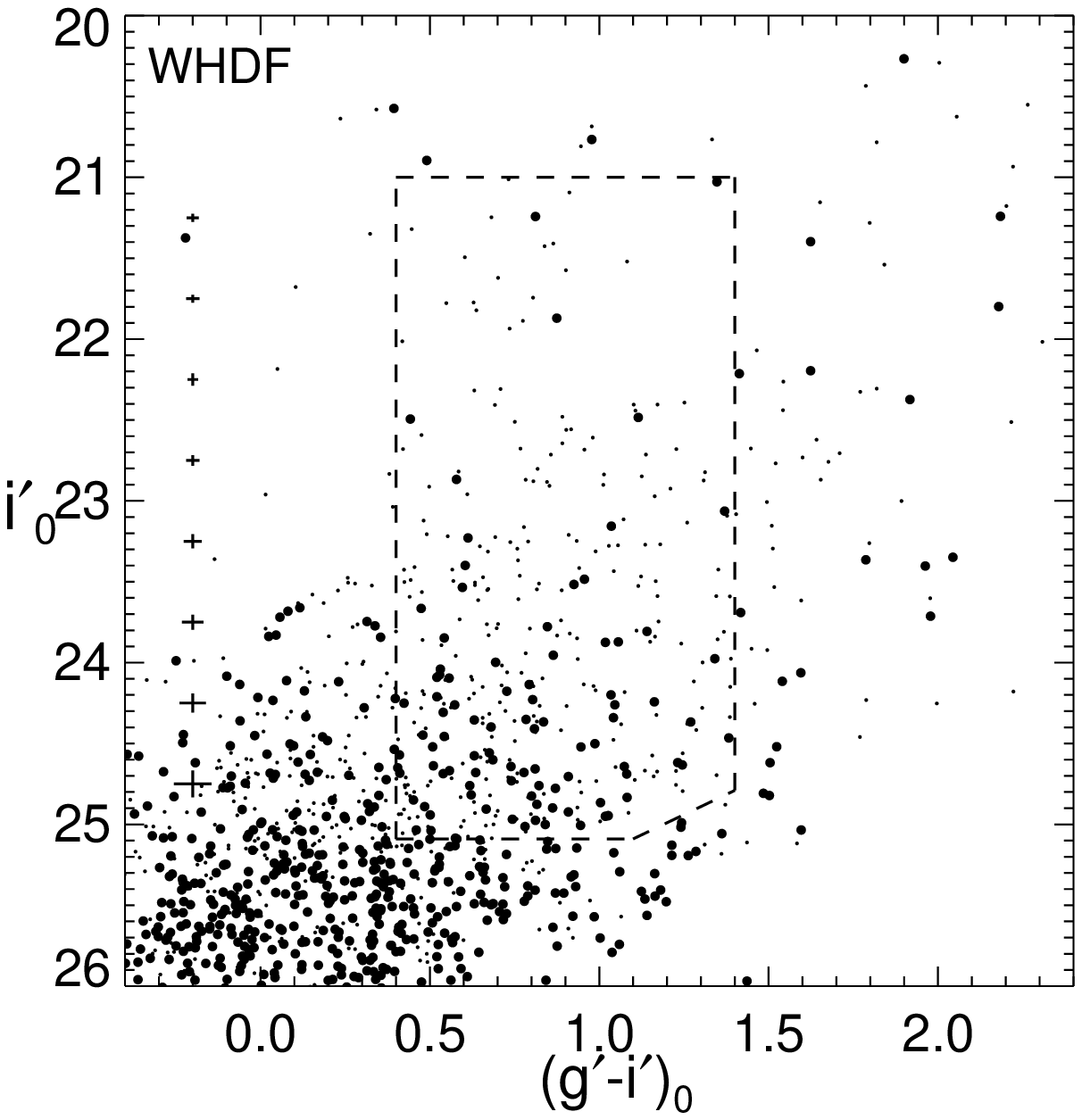}
\end{tabular}
\caption{Color-magnitude diagrams for sources detected around the five
  galaxies plus the control field (bottom right). Small black dots
  indicate the resolved sources, while the big black dots are the
  point sources as determined in Sect. \ref{sec:complete}. Globular
  clusters are depicted in red, inside the magnitude and color limits
  explained in the text. The number of GC candidates stated for each
  galaxy is {\it before} any decontamination and within the FOV. Green
  triangles in the NGC 3962 panel indicate UCD candidates (see Section
  \ref{sec:ucds}).}
\label{fig:cmds}
\end{figure*}

The reduction, photometry, artificial star experiments and
point-source definition of the WHDF were carried out in the same way
as the target galaxies, with the exception of the median filtering, in
the absence of a dominant galaxy. The color-magnitude diagram of
sources in the WHDF can be seen in Fig. \ref{fig:cmds} (bottom right
panel). Since the completeness of the WHDF is shallower than the
photometry in the NGC 2865 and NGC 4240 fields, the contamination in
the faint end of these fields was estimated by multiplying the number
of detected contaminant sources with the ratio of the interpolation
functions (Eq. \ref{eq:interp}) of the galaxies to the one of the
WHDF; generating bootstrapped samples with the expected number of
contaminant sources.

\subsection{SOAR/SAM imaging}
\label{sec:sam}

Additional imaging of NGC 3962 was acquired on the night of April 22,
2014, using the SOAR Adaptive Module (SAM) and its Imager (SAMI)
\citep{tokovinin10,tokovinin12}. SAM is a ground layer adaptive optics
system installed at the SOAR 4.1m telescope at Cerro Pach\'on, Chile.
SAMI gives a field of view (FOV) of $3\times3$ arcmin with a scale of
0.091\arcsec pixel$^{-1}$ ($2\times2$ binning).

In total, $5\times120$, $3\times120$ and $5\times120$ seconds
exposures were taken in the SDSS $g$, $r$ and $i$ filters,
respectively. Details of the reduction process will be given in a
forthcoming paper (Salinas et al. in prep), and only a preliminary
calibration to the standard system, not crucial for its scientific
goal, is given here. The FWHM measured on the combined images was
$0.59\arcsec$, $0.55\arcsec$ and $0.50\arcsec$ in the $g$, $r$ and $i$
filters, respectively. This set of images, of higher resolution than
the GMOS imaging, was used to study to the galaxy center where the
deeper GMOS images were saturated. More details are given in Sect.
\ref{sec:galaxies}.

\section{The GC color distribution}
\label{sec:color}

Color-magnitude diagrams for all the point sources in our target
galaxies can be seen in Fig. \ref{fig:cmds} as big black points, while
resolved sources are depicted will small black dots. Globular cluster
candidates were further limited to the color range $0.4<(g-i)_0<1.4$,
following previous GCS studies with the same filters
\citep{forbes04,wehner08,faifer11}, but with a slightly bluer limit to
allow the study of possible younger clusters \citep[as in NGC
  2685][]{sikkema06}. The bright limit was fixed at $M_i=-11.5$
following \citet{faifer11}, and the faint limit was given by the 80\%
completeness limit in both $g$ and $i$. These bounding limits
can be seen in Fig.  \ref{fig:cmds} with dashed lines. Photometry was
corrected for Galactic extinction using the \citet{schlafly11}
recalibration of \citet{schlegel98} dust maps.

 The effect of the color-dependent completeness is negligible for NGC
 3962 and the WHDF, mild for NGC 2771 and NGC 2865 and significant for
 NGC 4240. There is no color-dependent completeness for IC 4889.

None of the GCSs present an instantly obvious bimodal color
distribution as their more massive counterparts in cluster
environments \citep[e.g.][]{dirsch03}, but NGC 3962 stands out by its
healthy GCS. The GCSs with the deepest observations, NGC 4240 and NGC
2865, show a significant number of blue ($g'-i'\sim0.5$) sources below
$i'\sim25$, which are likely compact background galaxies which have
survived the point source determination process, while IC 4889 stands
out by the presence of a significant blue population throughout the
entire magnitude range ($21\lesssim i\lesssim24$), which seems absent
in the rest of the galaxies (compare, for example, with NGC 3962).

\begin{figure}
\begin{tabular}{c}
\includegraphics[width=0.98\linewidth]{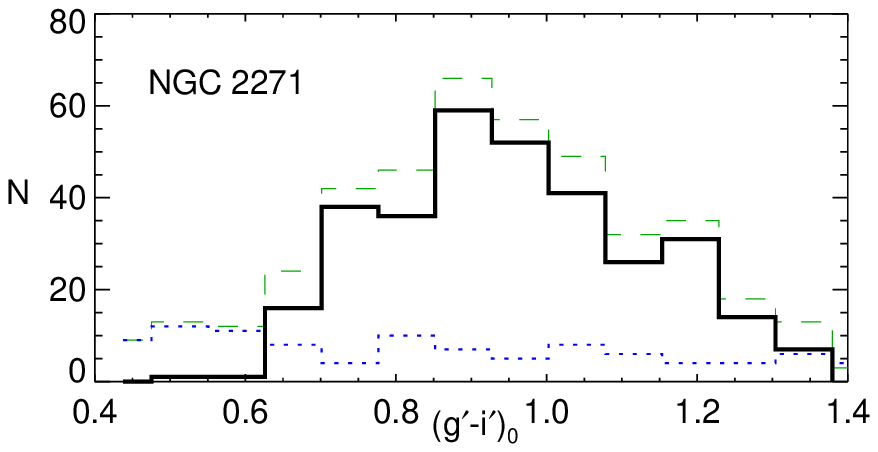}\\
\includegraphics[width=0.98\linewidth]{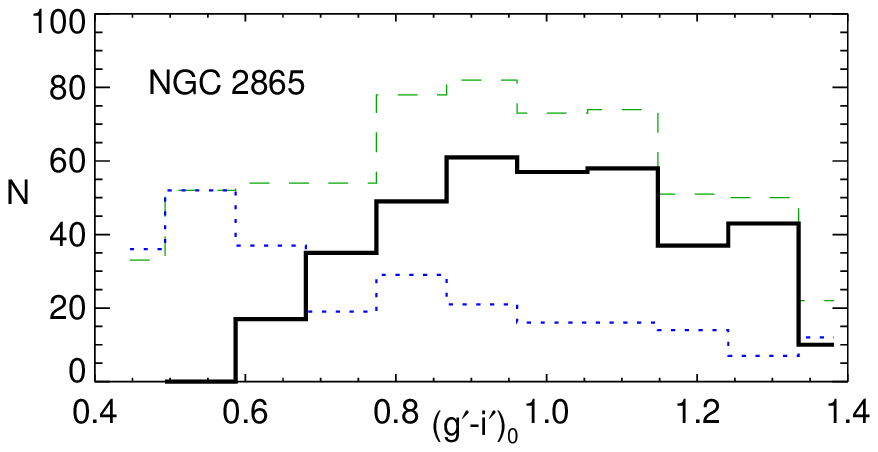}\\
\includegraphics[width=0.98\linewidth]{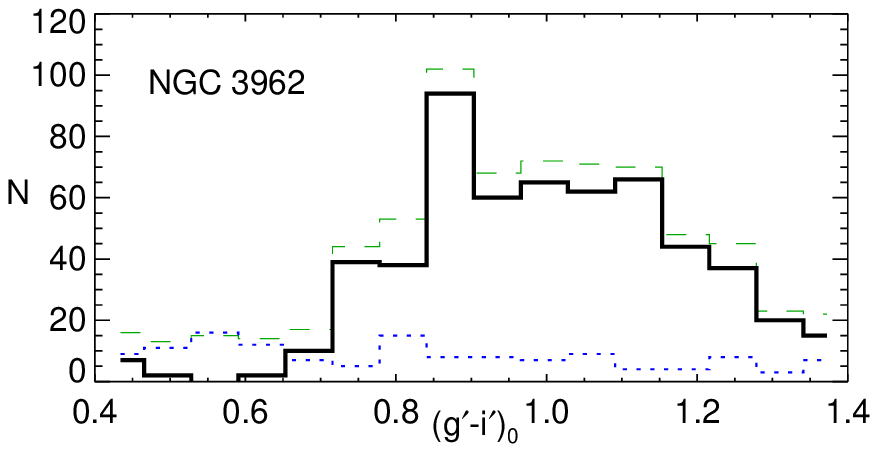}\\
\includegraphics[width=0.98\linewidth]{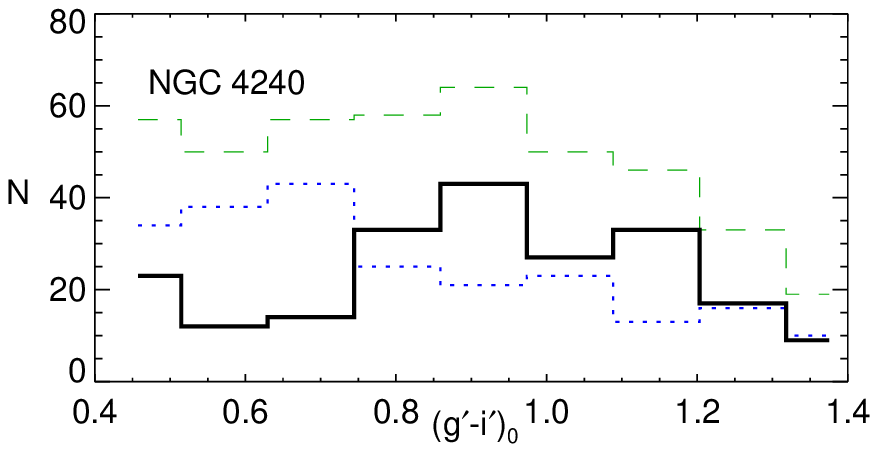}\\
\includegraphics[width=0.98\linewidth]{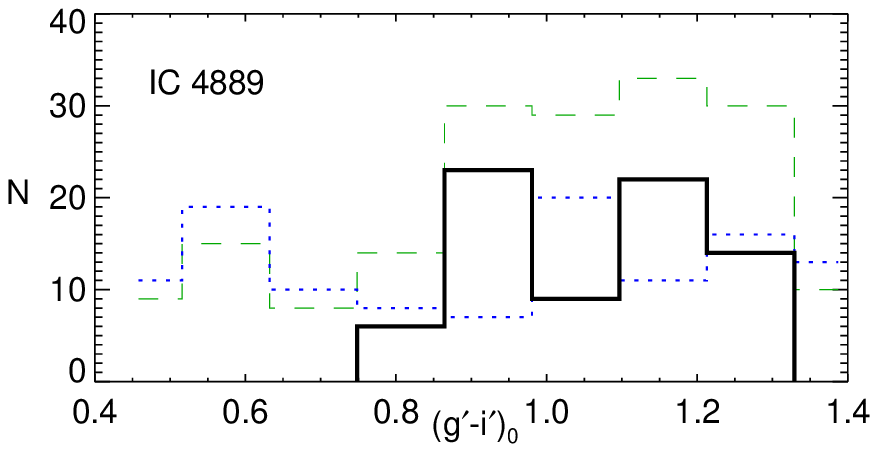}
\end{tabular}
\caption{Color histograms of the GCSs of the five target galaxies. The
  green dashed lines indicate the raw color distribution, the blue
  dotted line indicates the contamination in each field as determined
  in Sect. \ref{sec:contamination}. The thick black lines show the
  background-corrected distributions.}
\label{fig:histograms}
\end{figure}

Fig. \ref{fig:histograms} shows color histograms for the five GCSs,
where the data has been binned using an optimum bin size, which
depends on the sample size \citep{izenmann91}. Each panel shows the
raw (green dashed lines), background corrected (thick black lines) and
background estimate (blue dotted lines) color histogram for each GCS.
The background corrected histograms reveal that most of the GC
candidates with colors $(g'-i')_0\lesssim0.7$ can be accounted as part
of the contamination, except for NGC 2865 and especially NGC 4240
where a significant number of blue sources remain. The blue population
in NGC 2865 is consistent with the results of \citet{sikkema06} who
proposed the existence of a population of young clusters. Visually,
the clearest case for bimodality can be seen in the IC 4889 color
histogram, which contradicts the result of \citet{gebhardt99} which
found no evidence of bimodality in the GCS of this galaxy.

In order to test the existence of bimodality in the GC color
distributions more rigorously, we use a Gaussian mixture modeling
(\textsc{gmm}) as implemented by \citet{muratov10}. \textsc{gmm}
provides three indicators useful to discriminate between unimodal and
bimodal distributions: the kurtosis of the distribution, $k$, which if
negative indicates a flat-top distribution; the separation of the
peaks, $D$, defined as $D=|\mu_{\rm blue}-\mu_{\rm red}|/[(\sigma_{\rm
    blue}^2+\sigma_{\rm red}^2)/2]^{1/2}$, which, when larger than 2,
indicates a clear separation between the peaks \citep{ashman94}; and a
$p$--value, which gives the probability of obtaining the same $\chi^2$
from a unimodal distribution. Uncertainties in each statistic are
calculated with 100 bootstrapped realizations of the sample. To
further test the robustness of the results, \textsc{gmm} was run 50
times on samples generated by randomly subtracting the number of
contaminants expected per each color bin. The errors measured with
this method are always within the errors obtained from the
bootstrapped realizations (which are the quoted errors in Table
\ref{table:gmm}), but additionally provide an uncertainty for the
kurtosis which is not given by \textsc{gmm}. Given the large number of
blue sources in NGC 4240, \textsc{gmm} was run for sources with
$0.6\leq g-i \leq 1.4$, instead of $0.4\leq g-i \leq 1.4$ as the rest
of the galaxies.

The results of \textsc{gmm} applied over the background corrected GC
samples can be seen in Table \ref{table:gmm}. Following
\citet{usher12} we define a galaxy as showing clear bimodality when
the conditions p-val$<0.1$, $D>2$ and $\kappa<0$ are simultaneously
met. These criteria are clearly fulfilled by NGC 2271, NGC 2865 and IC
4889. NGC 4240 also fulfill the criteria, although the large 
number of blue sources makes its case less reliable.

The most interesting case is given by NGC 3962. Despite being the
galaxy in our sample with the richest GCS and with deep observations,
its peak separation is barely larger than 2, the kurtosis is very
close to zero and has the highest p-value, making the case for
bimodality very doubtful. Intermediate-age globular clusters are known
to smear out bimodality \citep[e.g.][]{richtler12}, but the luminosity
function shows no sign of intermediate-age clusters (see
Sect. \ref{sec:lumfun}). Only observations using a longer wavelength
baseline will provide a more definitive answer.

Additionally, Table \ref{table:gmm} gives the red fraction, $f_r$, for
each GCS. Red fractions, simply defined as the number of red clusters
over the total number of clusters, are further discussed in
Sect. \ref{sec:red_fraction}.

\begin{table}
  \caption{Globular cluster radial profiles.}
  \label{table:surface}
 \centering
  \begin{tabular}{@{}lrc@{}}
  \hline\hline
Galaxy   &   \multicolumn{2}{c}{power-law fit}\\

 & \multicolumn{1}{c}{$\sigma_{0}$} & $n$ \\
\hline
 NGC 2271 & $1.69\pm0.35$&$-2.18\pm0.18$\\
 NGC 2865 & $1.38\pm0.29$&$-1.88\pm0.15$\\
 NGC 3962 & $1.29\pm0.27$&$-1.81\pm0.14$\\
 NGC 4240 & $ -0.77\pm0.30$&$-0.89\pm0.15$\\	
 IC 4889  & $-1.91\pm0.84$&$-0.61\pm0.43$\\
\hline
\end{tabular}
\end{table}

\begin{table*}
\tiny
\caption{\textsc{gmm} results.}
\label{table:gmm}
 \centering
  \begin{tabular}{@{}lccccccccccccc@{}}
    \hline\hline
    Galaxy & $N$&$\mu_{\rm blue}$ & $\mu_{\rm red}$& $\sigma_{\rm blue}$& $\sigma_{\rm red}$& $f_r$& $D$&$\kappa$&p-val\\
    \hline
\noalign{\smallskip}
    NGC 2271&295&$0.896\pm0.014$&$1.184\pm0.025$&$0.125\pm0.011$&$0.067\pm0.011$&$0.17\pm0.06$&$2.87\pm0.26$&$-0.43\pm0.04$&0.001\\
    NGC 
2865&332&$0.848\pm0.043$&$1.125\pm0.044$&$0.128\pm0.019$&$0.096\pm0.02
0$&$0.38\pm0.14$&$2.43\pm0.19$&$-0.87\pm0.03$&0.001\\
    NGC 
3962&557&$0.947\pm0.057$&$1.208\pm0.041$&$0.156\pm0.019$&$0.089\pm0.01
9$&$0.19\pm0.12$&$2.05\pm0.37$&$-0.11\pm0.04$&0.007\\
    NGC 
4240&123&$0.765\pm0.055$&$1.098\pm0.063$&$0.108\pm0.033$&$0.103\pm0.03
3 $&$0.44\pm0.17$&$3.08\pm0.35$&$-1.16\pm0.04$&0.001\\
    IC  4889&74 
&$0.922\pm0.015$&$1.187\pm0.019$&$0.062\pm0.010$&$0.081\pm0.010$&$0.54
\pm0.07$&$3.65\pm0.43$&$-1.25\pm0.05$&0.001\\
    \hline
\end{tabular}
\tablefoot{The second column indicates the number of (background subtracted) GC candidates used as input for GMM. Columns 3 and 4 indicate the peak color of the blue and red subpopulations, while columns 5 and 6 indicate the dispersion of the best fitting Gaussians. Column 7 indicates the red fraction, that is, the number of cluster associated with the red subpopulation divided by the total number of GC candidates. Column 8 is the peak separation as defined in Sect. \ref{sec:color}, while column 9 is the sample kurtosis.}
\end{table*}

\section{The GC radial distribution}
\label{sec:radial}

\begin{table*}
\tiny
\centering
\caption{GCLF fitted parameters and specific frequencies.}
\label{table:gclf}
 \centering
  \begin{tabular}{@{}lcccccccc@{}}
  \hline\hline
Galaxy & $a_0$            &$m_0$           & $\sigma_{m}$ & $N_{GC}$    & $S_{N}$ &    $T_{N}$ &$      T_{\rm blue}$&   $T_{\rm red}$\\
\hline
NGC 2271 &$84   \pm 1$   &$24.54$\,(f)    &$1.16\pm0.02$& $562\pm9$  
&$1.46\pm0.38$& $5.87\pm0.10$&$4.87\pm0.25$&$1.00\pm0.35$ \\
NGC 2865 &$61   \pm 2$   &$24.92$\,(f)    &$1.02\pm 1.1$& $410\pm8$  
&$1.00\pm0.19$& $2.51\pm0.05$&$1.55\pm0.48$&$0.95\pm0.35$ \\
NGC 3962 &$128  \pm 15 $ &$24.50\pm0.21$  &$1.44\pm0.15$& $854\pm98$ 
&$1.26\pm0.57$& $2.87\pm0.32$&$2.33\pm0.39$&$0.55\pm0.35$ \\
NGC 4240 & --            & --             & --          & $84 \pm9$  
&$2.15\pm0.80$& $2.53\pm0.27$&$1.42\pm0.55$&$1.11\pm0.45$ \\
IC 4889  & --            & --             & --          & $280\pm17$ &$0.85\pm0.12$& $2.55\pm0.15$&$1.17\pm0.19$&$1.38\pm0.20$ \\
\hline
\end{tabular}
\end{table*}

\begin{figure}[]
\begin{tabular}{c}
\includegraphics[width=0.98\linewidth]{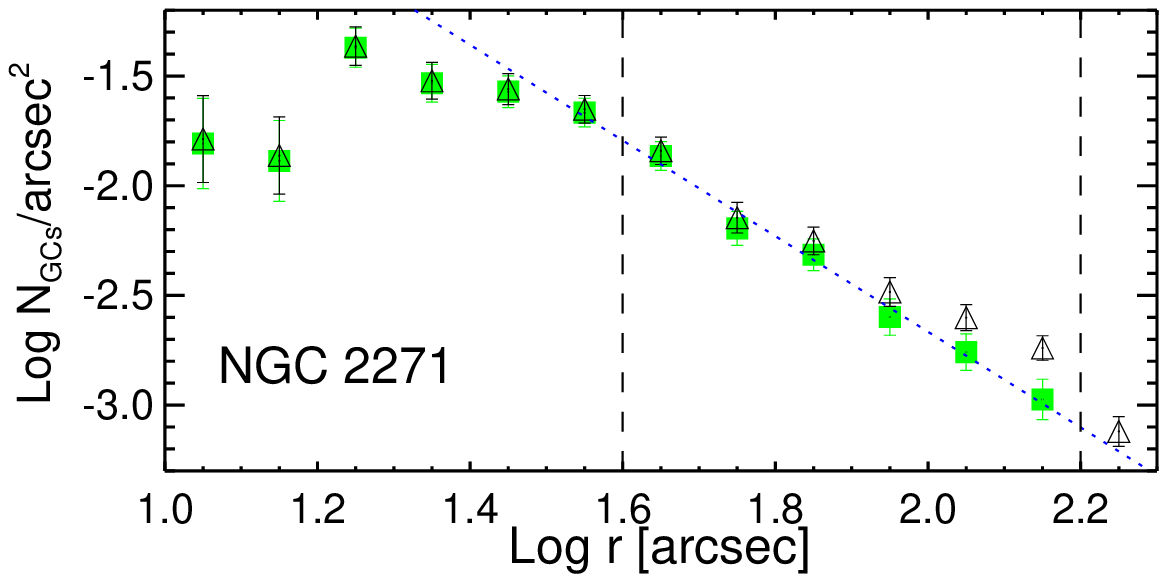}\\
\includegraphics[width=0.98\linewidth]{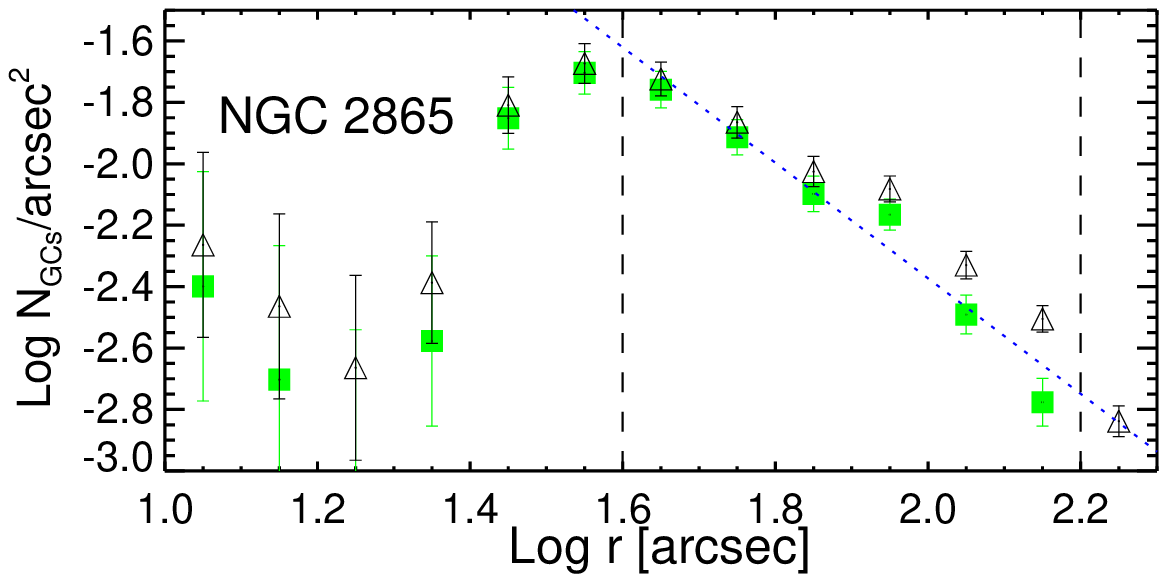}\\
\includegraphics[width=0.98\linewidth]{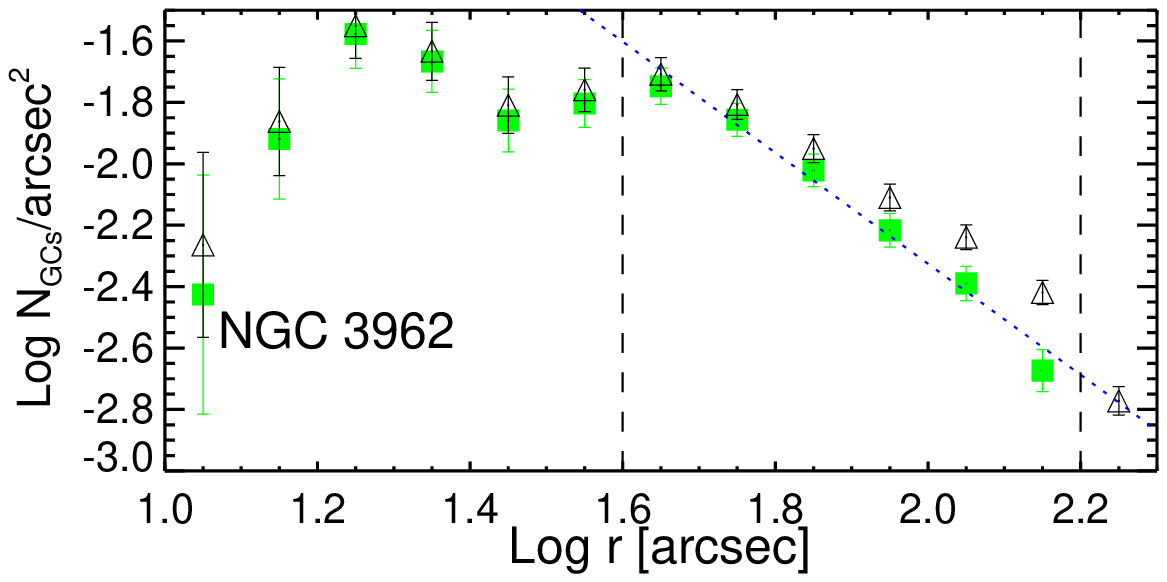}\\
\includegraphics[width=0.98\linewidth]{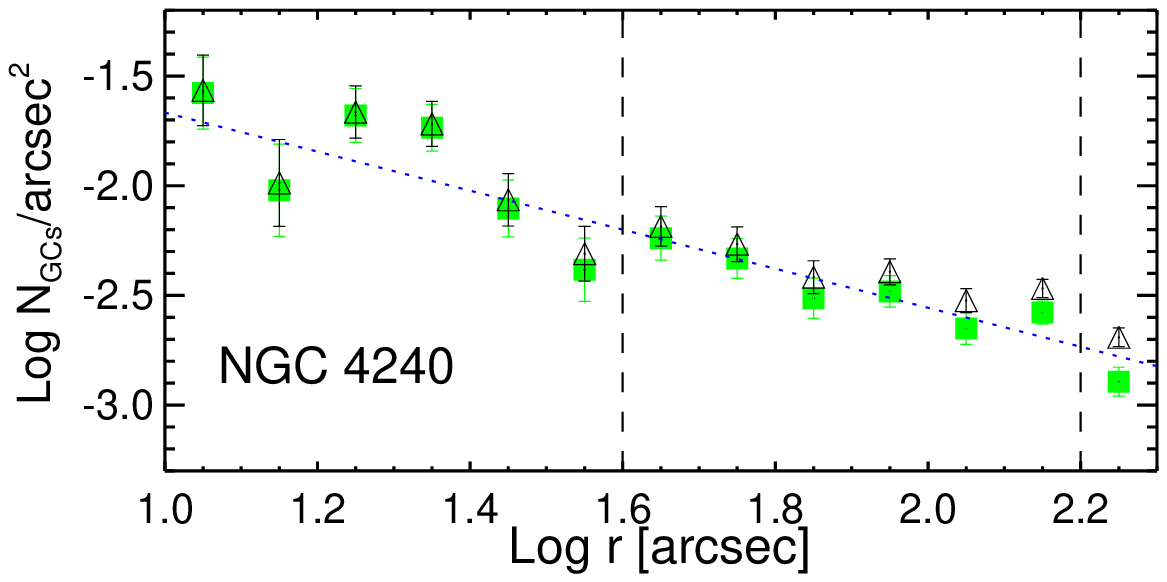}\\
\includegraphics[width=0.98\linewidth]{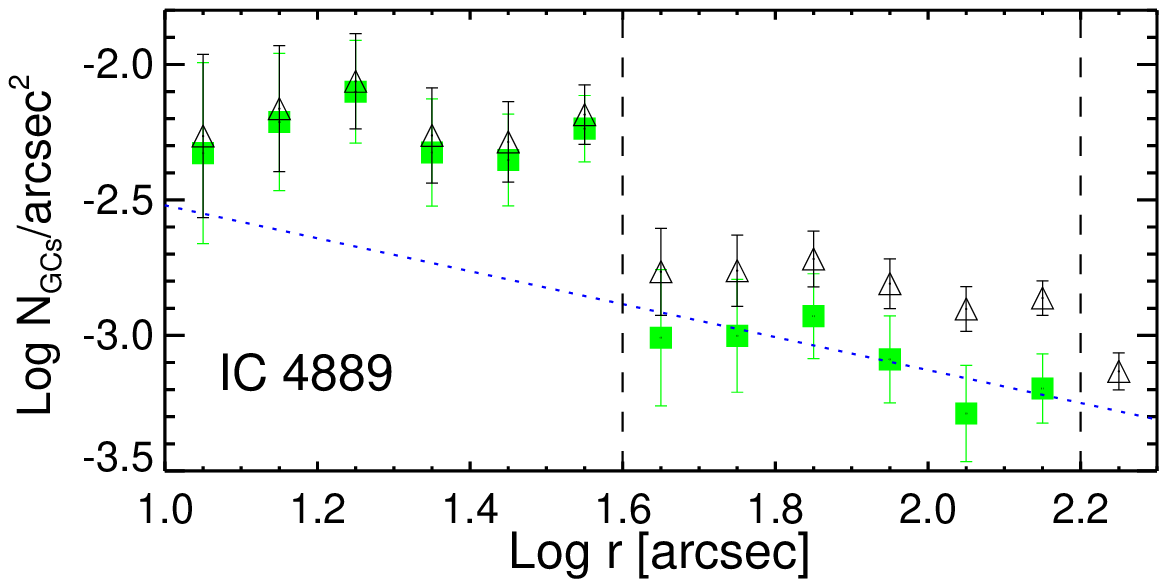}
\end{tabular}
\caption{The radial distribution of globular clusters in the studied
  galaxies. Open black triangles indicate the total population, while
  green filled squares depict the background subtracted
  population. Each distribution was fitted with a power law (blue
  dotted line), between the limits indicated by horizontal black
  dashed lines.}
\label{fig:radial}
\end{figure}

For photometry obtained with wide enough FOVs, the population of 
GCs detected around a galaxy is expected to drop off to a constant 
value at sufficiently extended radial positions from the galactic
center. This constant number will represent foreground stars and
faint background galaxies and the projected surface density of GCs as
a function of galactocentric distance will give the radial
distribution of GC candidates. In practice, given the rather limited 
FOV, we adopted the outermost bin of the radial distribution 
as background. Starting from the center of the images
and considering only detected GC candidates brighter than the 80\%
completeness magnitude in the $i$ filter, we grouped GC candidates and
contaminants into concentric circular annuli, subtracted the 
contamination in each annulus and then divided by the effective area 
over which GC candidates are distributed.

We fitted power laws, $\sigma_{\rm GC}=\sigma_{0}r^{n}$, to the 
surface density profile of galaxies in our sample, where $\sigma_{\rm 
GC}$ is the number of GCs per square arc second and $r$ is the 
projected radial distance. The fit was done for all GCSs in the 
fixed interval $40\arcsec<r<160\arcsec$ (horizontal dashed line in 
Fig. \ref{fig:radial}), which avoids the central zone of 
incompleteness produced by saturation, ill-subtracted galaxy and 
possible dusty features (see Sect. \ref{sec:galaxies}). Results of 
the fit can be seen as dotted blue lines in Fig. \ref{fig:radial} and 
in Table \ref{table:surface}.

The richest GCS, NGC3962, NGC 2271 and NGC 2865 follow a power law 
with index $n\sim-2$, as is the case for most giant ellipticals 
\citep[e.g.][]{bassino06}. The galaxy which presents the larger 
deviation from the power law in the selected radial range is NGC 
2865, which shows a small overabundance of GCs at $\log 
r\sim1.95$ (90$\arcsec$) which might be associated with its 
intricate shell system.

NGC 4240 and IC 4889 present very shallow GC profiles, probably an 
effect of the low number of GCs present in these systems. 
Particularly problematic is the case of IC 4889 where an abrupt jump 
in the cluster distribution is visible at $r\sim40\arcsec$. IC 4889 
not only possesses a poor cluster system, but also is the galaxy with 
the shallowest observations of the set. 

Finally, we note that the 
galaxies with fewer dusty features (NGC 2271 and NGC 4240, see 
Section \ref{sec:color}) are the ones which follow closer the power 
law distribution inside the fit limit of $40\arcsec$.

\begin{figure}
\begin{tabular}{c}
\includegraphics[width=0.98\linewidth]{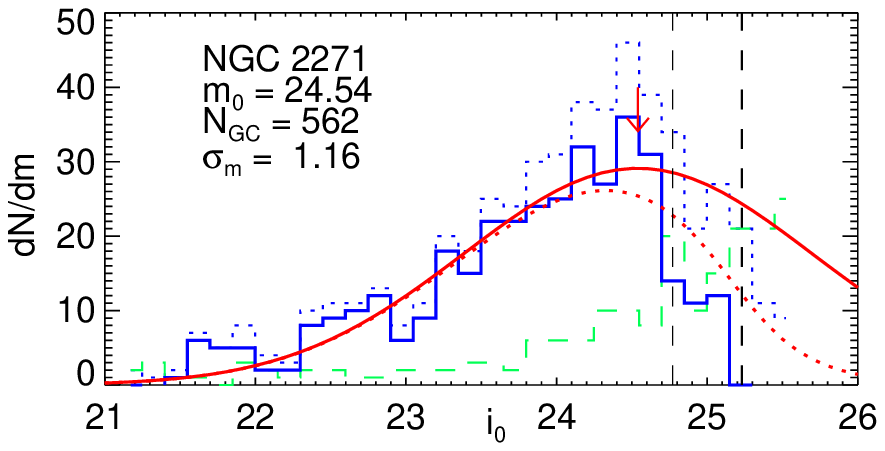}\\
\includegraphics[width=0.98\linewidth]{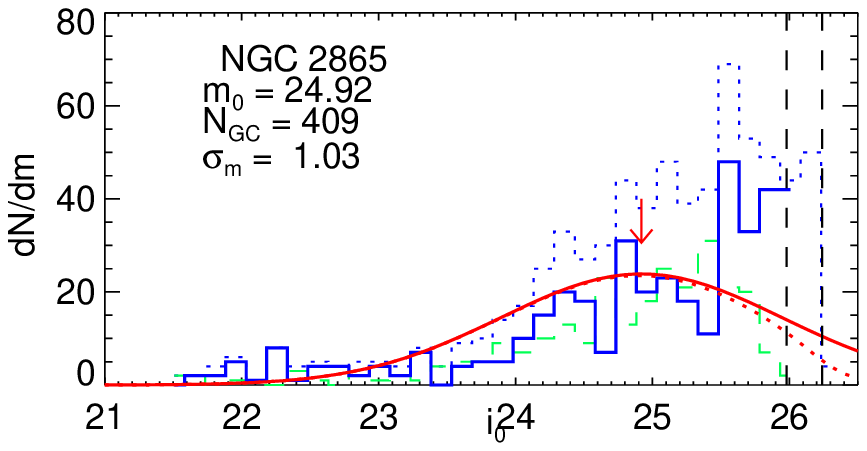}\\
\includegraphics[width=0.98\linewidth]{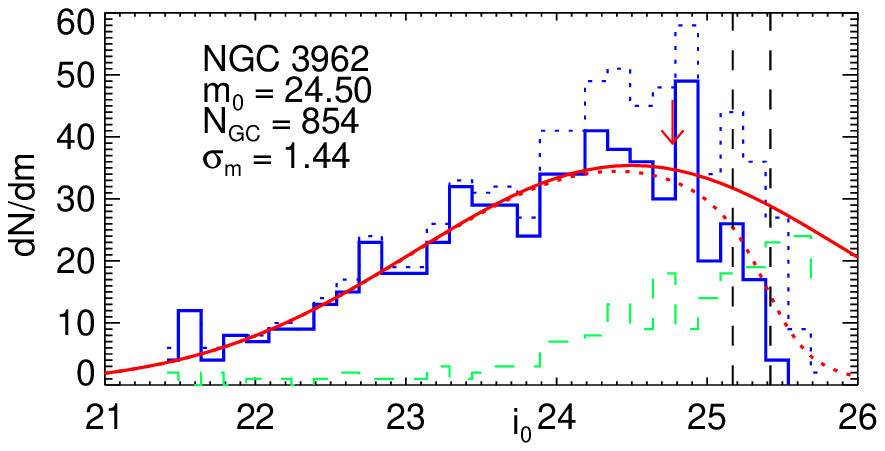}\\
\includegraphics[width=0.98\linewidth]{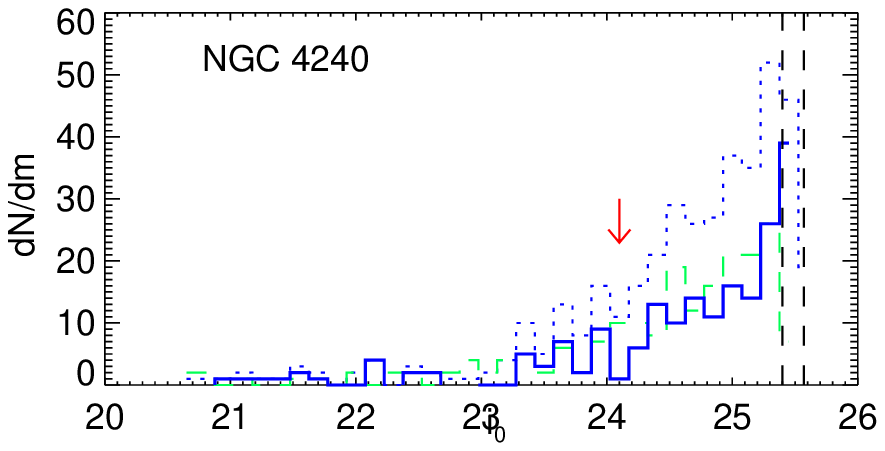}\\
\includegraphics[width=0.98\linewidth]{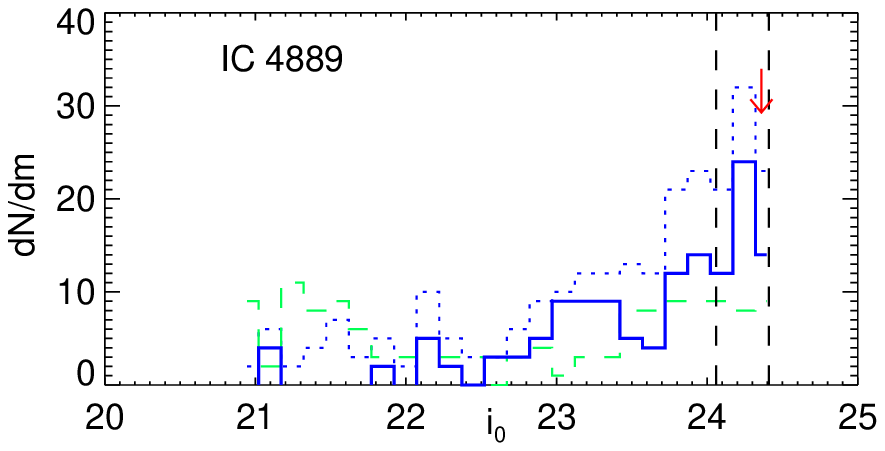}\\
\end{tabular}
\caption{GCLF for the five GCSs. Blue lines indicate the raw GCLF
  (dotted), the LF of the contamination (dashed), and the
  contamination-corrected GCLF (solid). Red lines indicate the fitted
  function (Eq. \ref{eq:gclf_eqn}, dotted line) and its respective
  Gaussian. Vertical dashed black lines indicate the 80\% and 50\%
  completeness limits in $i'$ as obtained in
  Sect. \ref{sec:complete}.}
\label{fig:lum_func}
\end{figure}

\section{Luminosity functions and distances}
\label{sec:lumfun}

Luminosity functions in the $i'$ band for the point sources in each
galaxy field can be seen in Fig. \ref{fig:lum_func} (blue dotted
lines). Background/foreground contamination was estimated using the
WHDF and Besan\c{c}on models as in Sect. 
\ref{sec:contamination} (green dashes lines). The 
background-corrected luminosity
functions, that is, the ones for the globular cluster candidates,
are depicted with solid blue lines.

The inverted red arrows indicate the expected position of the turnover
magnitud (TOM) of the globular cluster luminosity functions (GCLFs),
assuming a universal value of $M_I^{\rm TOM}=-8.46$ \citep{kundu01},
transformed to $M_i$ using Eq. 2 from \citet{faifer11}, and distance
moduli as given in Table \ref{table:sample}.

For NGC 2271, NGC 2865 and NGC 3962, the contamination-subtracted
GCLFs were modelled as the product of a Gaussian function with the
completeness function (Eq. \ref{eq:interp}):
\begin{equation}
\label{eq:gclf_eqn}
\frac{\ud N}{\ud m}=\frac{a_0}{\sqrt{2\pi}\sigma_m}\exp{\left[\frac{(m-m_0)^2}{2\sigma_m^2}\right]}^{\frac{1}{2}}\left[1-\frac{\alpha(m-m_{50})}{\sqrt{1+\alpha^2(m-m_{50})^2}}\right],
\end{equation}
where $m_0$ is the TOM and $\sigma_m$ the width of the GCLF. The
parameters for the completeness interpolation function are the ones
derived in Sect. \ref{sec:complete}. The fitting of equation
\ref{eq:gclf_eqn} was done using \textsc{mpfit}, an implementation of
non-linear square fitting in IDL \citep{markwardt09}. The luminosity
functions were fitted down to the 50\% completeness detection limit in
$i'$ (Table \ref{table:complete}), with the exception of NGC 2865 were
the fit was done using magnitudes down to the 80\% limit, given the
large number of sources at $i>25.5$, which are probably faint
background galaxies that escaped the point-source criteria defined in
Sect. \ref{sec:obs}.

NGC 3962 is the only galaxy where the completeness limit is well
beyond the TOM, therefore only for this galaxy we fit $m_0$,
$\sigma_m$ and the Gaussian amplitude, $a_0$. In the cases of NGC 2271
and NGC 2865, the TOM was fixed to its expected value based on its SBF
distance \citep{tonry01}, and only $\sigma_m$ and $a_0$ were fit. The
results of the fit for these three galaxies can be seen in Table
\ref{table:gclf}.

Only the case of NGC 3962 provides material to discuss its distance
based on its GCLF. The distance modulus derived from the TOM position
is $32.47\pm0.20$, while the \citet{tonry01} value, assumed throughout
this paper, and based on SBF measurements is $32.74\pm0.40$. Even
though they agree within the erros, the GCLF based distance is
significantly lower than the SBF distance. If the SBF distance were
the correct one, this would imply a slightly brighter TOM,
TOM$_i=-8.24$, instead of the expected $-7.97$.

As discussed in Sect. \ref{sec:color}, a possible explanation for the
absence of a clear separation between the color peaks in NGC 3962 is
the presence of an intermediate age population, but intermediate age
populations have exactly the opposite effect, producing fainter TOMs
\citep[e.g.][]{richtler03}, also even though metallicity is expected
to have an influence on the TOM, this is minimized when using $I$ band
observations \citep[e.g.][]{rejkuba12}. A brighter TOM agrees with
\citet{blakeslee96} who found evidence that the TOM varies with
environment, with low-density galaxies having a brighter TOM
\citep[see also][]{villegas10}. A larger sample of IEs with deep
enough observations would be necessary to confirm this.

The radial profiles in Fig. \ref{fig:radial} indicate that the density
of GCs quickly drops to values close to zero within the FOV,
therefore, we did not attempt an extrapolation of the total number of
clusters, $N_{GC}$, to obtain the fraction outside the FOV. The total
number of GCs is then simply derived from the Gaussian fitted to the
GCLF. In the cases where no fitting was attempted (NGC 4240 and
IC4889), the total number of GCs was estimated by doubling the number
of detected clusters above the TOM. The GCLF in NGC 4240 differs 
from the expected shape close to a Gaussian, with a large 
number of very faint sources. Given the old age of the galaxy (Sect. 
\ref{sec:sample}) and the lack of evidence from the color maps of a 
recent meger event (Sect. \ref{sec:galaxies}), these are probably a 
residual contamination from background galaxies and not a faint 
young population of clusters. A similar GCLF is seen in NGC 3377 as 
shown by \citet{cho12}.

Total number of GCs for the five
galaxies can be seen in Table \ref{table:gclf}.

\section{Galaxies surface brightness and colors}
\label{sec:galaxies}

Surface brightness profiles of the galaxies were measured via ellipse
fitting using the \textsc{iraf} task \textsc{ellipse}. All the other
objects in each galaxy field were initially masked using the
segmentation image produced by \textsc{SExtractor}, but additional
regions (extended halos of bright stars, chip defects, etc) were
interactively masked during the \textsc{ellipse} execution. Once the
regions to mask were established, \textsc{ellipse} was run a second
time, fixing the center of the ellipses as the mean central value of
the ellipses with semi-major axes between 30 and 300 pixels. Sky
brightness was calculated using the unmasked pixels outside the last
measured isophote, taking the robust mean \citep{beers90} of 40 random
samples containing 800 sky pixels each. The uncertainty in the sky
value was taken as the dispersion of these 40 sky
measurements. Photometric calibration was done with the same equations
as the point source photometry (Table \ref{table:coeff}), correcting
for Galactic extinction following \citet{schlafly11} dust maps.

The results from the ellipse fitting can be seen in Appendix
\ref{sec:surface}. The errors in the $g'$ and $i'$ surface brightness
profiles comes from the sum in quadrature of the rms scatter in the
intensity from the ellipse fitting and the uncertainty on the sky
value. Surface brightness is given as function of the geometric mean
radius, $r=a\sqrt{1-\epsilon}$, where $a$ is the fitted ellipse major
axis and $\epsilon$ its ellipticity.

Color maps were also generated for each galaxy in order to look for
color anomalies related to dust and possible past
mergers. 

NGC 2271 shows a flat color profile in the inner $\sim 30 
\arcsec$ along the major axis, with noticeable color gradient at 
larger radius (Fig. \ref{fig:colorn2271}). This is the 
normal behaviour for early-type galaxies \citep[e.g.][]{labarbera12}, 
and compatible with the negligible age and metalicity gradients found 
by \citet{reda07} inside $0.5R_e$. No dusty features are visible.

Fig. \ref{fig:colorn2865} (left panel), shows the color map of NGC 
2865. NGC 2865 is widely considered as a merger remnant 
\citep{hau99}. The right panel shows the galaxy surface brightness 
map with the same scale. The large of number of shells, bluer than 
the galaxy body, have been discussed elsewhere 
\citep{malin83,fort86}. We just emphasize a very narrow $\sim 20 
\arcsec$ long and $\sim 1.5\arcsec$ wide stream (white box in 
Fig. \ref{fig:colorn2865}), within the structure labeled  ``jet'' in 
\citet{fort86} (their Fig. 3), pointing in projection directly to the 
galaxy center. Its blue color together with the large shell at the 
exact opposite side of the galaxy, probably makes it part of the 
same accretion process that produced the shells, although these very 
narrow radial features are not reproduced by simulations 
\citep[e.g.][]{hernquist87,bilek14}, which show rather fan-like 
structures.

Fig. \ref{fig:colorn3962} shows the $g-i$ color map for NGC
3962. The same structures detected by \citet{buson93} using H$\alpha$
imaging can be seen. First, an extended, slightly off-centered
arc-like structure about $15\arcsec$ from the galaxy center, with
redder color than the underlying galaxy light. The SAM images
(Fig. \ref{fig:colorn3962}, right panel), also show a central
elongated component, misaligned with the major and minor axis of the
galaxy. Ionized gas is now commonly thought to be of external origin,
especially for massive, round ellipticals \citep[e.g.][]{sarzi06}, so
even though the large scale structure of the galaxy does not reveal
any evidence for past accretion, as indicated by the very low tidal
parameter measured by \citet{tal09}, the inner color structure might
be indication of an older merger.

NGC 4240 shows a flatter color profile (Fig. \ref{fig:colorn4240}),
similar to the one measured on the field elliptical NGC 7507
\citep{salinas12}. This agrees with the shallow age and metallicity
gradients found by \citet{reda07}. A notable feature is the $\Delta
(g-i) \sim 0.15$ color difference, for the inner $\sim 15\arcsec$,
between the South and North sections. Even though one would be tempted
to relate it to the lop-sided rotation curve found by \citet{hau06} on
the same scales, that rotation pattern was found roughly along the
East-West axis. Another feature is a red plume of $\sim 15 \arcsec$ in
the SE direction (indicated with a black arrow in
Fig.\ref{fig:colorn4240}).  Both features were undetected by
\citet{reda04}.

IC 4889 shows a large spiral dusty feature (Fig. 
\ref{fig:coloric4889}), very similar to the one on NGC 3962. A 
remarkable feature is the unphysically red external halo. Despite our 
efforts to understand any issues with these images, we were unable to 
pinpoint the origin for the very flat $i$ profile.

\section{Discussion}
\label{sec:discussion}
\subsection{The specific frequency of clusters in low- and high-density environments}

The globular cluster specific frequency, $S_{N}$, defined by
\citet{harris81} as
\begin{equation}\label{eq:sn}
  S_{N}=N_{GC}10^{0.4(M_{V}+15)}, 
\end{equation}
connects the luminosity of a galaxy with its globular cluster
population, in an attempt to measure how efficient is globular cluster
formation compared to star formation. In Eq. \ref{eq:sn}, $N_{GC}$ is
the total number of globular clusters in a galaxy, and $M_{V}$ is its
absolute $V$ magnitude.

\begin{figure}
\includegraphics[width=0.48\textwidth]{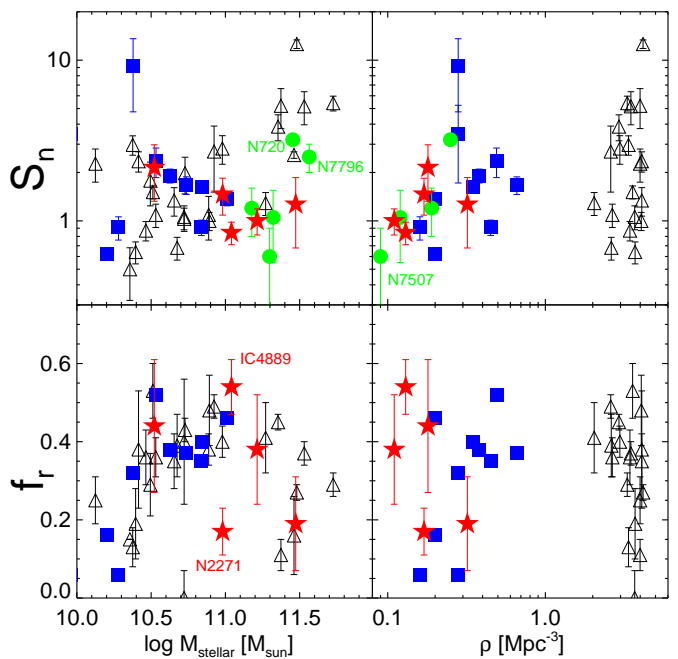}
\caption{Specific frequency (upper panels) and globular cluster red
  fraction (lower panels) as function of stellar mass and the
  \citet{tully88} density parameter. Red symbols represent our sample,
  while blue squares are taken from the \citet{cho12} sample. Black
  open triangles are the results from the ACSVCS \citep{peng08}, while
  the the green filled circles are five more IEs described in the
  text.}
\label{fig:sn}
\end{figure}

We obtain $S_{N}$ values for our sample using $N_{GC}$ as obtained
from the GCLF analysis in Section \ref{sec:lumfun}. Figure
\ref{fig:sn} (upper panel) contains a comparison between our results
(red stars) with the \citet{cho12} sample of 10 low-density early-type
galaxies (blue squares) and galaxies from the ACS Virgo Cluster Survey
\citep[ACSVCS, open black triangles,][]{peng08}. Additionally, we
include the $S_N$ of five more isolated/low-density ellipticals, NGC
3585, NGC 5812 \citep{lane13}, NGC 7507 \citep{caso13}, NGC 720 
\citep{kartha14} and NGC 7796 \citep{richtler15} (green filled 
circles).

Fig. \ref{fig:sn} (upper left panel) presents $S_N$ as a function of
stellar mass. Stellar masses for each galaxy in our sample and
\citet{cho12} sample were derived using homogeneous $K_s$ apparent
magnitudes taken from the 2MASS Extended Source Catalog
\citep{jarrett03}. $K$-band luminosities were transformed into stellar
masses by using the Hubble type dependent mass-to-light ratios from
\citet{spitler08}, based on \citet{bruzual03} population
models. Stellar masses for the ACSVCS galaxies were taken directly
from \citet{peng08}. Distances for the ACSVCS galaxies come from the
recalibrated surface brightness fluctuations measurements given by
\citet{blakeslee09}.

Isolated ellipticals have a low $S_N$, scattered around $S_N\sim1.5$,
although there is a trend in the $10.5<\log M_{\star}<11.4$ in the
sense that less massive galaxies have $S_N$ closer to 2, while the
more massive are clustered around $\sim1$. A Spearman's rank
correlation test gives a coefficient of $-0.74$, with a two-sided
significance of 0.002, indicating a high correlation. Considering a
similar mass range, the opposite trend is seen for the Virgo GCSs
\citep[already discussed by][]{peng08}. IEs with $\log 
M_{\star}>11.4$ show a diverse behavior, while NGC 3962 follows the 
trend of less massive galaxies, NGC 720 
\citep[$S_N=3\pm0.2$,][]{kartha14} and NGC 
7796 \citep[$S_N=2.5\pm0.5$,][]{richtler15} show an increased $S_N$, 
although still lower than Virgo ellipticals of similar mass.

\begin{figure}
\includegraphics[width=0.48\textwidth]{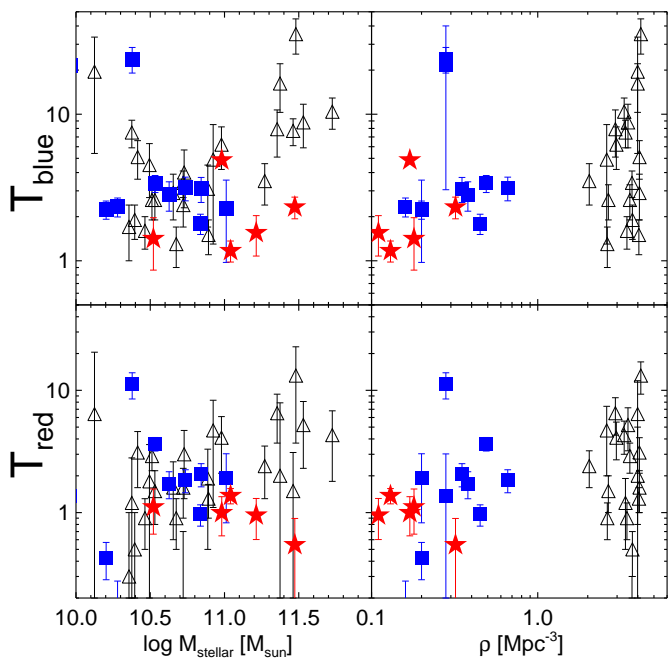}
\caption{$T$-parameters (see text) as function of stellar mass and
  density parameter. Symbols are the same as in Fig.\ref{fig:sn}.}
\label{fig:T}
\end{figure}

The upper right panel of Fig. \ref{fig:sn} shows now the $S_N$ as a
function of \citet{tully88} density parameter. Galaxies in both
low-density and high density environments share a broad range of
$S_N$, but again a trend within the low-density sample can be seen,
where the most isolated galaxies ($\rho\sim0.1$) have even lower
$S_N$, altough this trend almost totally driven by NGC 7507, the
galaxy with the lowest density in the sample and with a surprisingly
low $S_N$ \citep{caso13}. Note that several galaxies in Virgo lack a
density parameter, as well as NGC 4240 and NGC 7796, which are not
included in the \citet{tully88} catalog due to their distances.

The $S_N$ depends on $M_V$, which in turns depends strongly on the 
age. Given that isolated ellipticals are considered late accretors, 
spanning a wide range of ages \citep[e.g.][]{kuntschner02,reda07}, a 
true comparison with cluster ellipticals, believed to be 
homogeneously old, is hampered. To compare galaxies with possible 
different stellar populations, it is relevant to use the $T_N$ 
parameter introduced by \citet{zepf93},

\begin{equation}
  T_N=N_{GC}/(M_{\star}/10^9) ,
\end{equation}
where the number of GCs is normalized to the stellar mass,
$M_{\star}$, instead of $M_V$. Additionally, the parameters $T_{\rm
  blue}$ and $T_{\rm red}$ can be defined, where the number of blue
and red clusters are used instead of the total number, $N_{GC}$. The
amount of blue and red clusters are obtained using the red fractions
measured in Sect. \ref{sec:color}.

Figure \ref{fig:T} shows $T_{\rm blue}$ and $T_{\rm red}$ for our
sample together with the same galaxies used in the $S_N$ comparison;
the \citet{cho12} sample of low-density environment early-type
galaxies, and early-type galaxies from the ACSVCS
\citep{peng06,peng08}. Colors and symbols are the same as in
Fig. \ref{fig:sn}.

The trend with mass seen for the $S_N$ is not seen for $T_{\rm blue}$,
although the more massive isolated ellipticals still show
significantly lower values when compared to Virgo ellipticals of the
same mass. The $T_{\rm red}$ values for our sample are systematically
lower than the \citet{cho12} and ACSVCS samples in the range of masses
described.

Finally, the right panels of Fig. \ref{fig:T} shows the behavior of
the $T_{\rm red}$ and $T_{\rm blue}$ parameters as function of the
density parameter. The trend seen for $S_N$ among isolated and
low-density galaxies is repeated for $T_ {\rm blue}$ although less
clearly. No clear trend is seen for $T_{\rm red}$, although if we
solely considering our sample without the \citet{cho12} galaxies is
clear than the $T_{\rm red}$ values for isolated ellitpicals span a
narrow range close to 1 or less, while Virgo ellipticals have on
average higher values.

\subsection{The red fractions of isolated ellipticals}
\label{sec:red_fraction}

The proportion between blue and red clusters in a GCS is usually
measured using the red fraction of GCs \citep[e.g.][]{peng06}, that
is, the fraction of red clusters compared to the entire population in
a GCS. Red fractions are an important tool to discriminate between the
possible scenarios of GCS formation.

Red fractions for our galaxies derived in Sect. \ref{sec:color} are
presented in Fig. \ref{fig:sn} (lower panels, red stars), together
with red fractions for GCSs belonging to the ACSVCS \citep[][black
triangles]{peng08} and the 10 low-density ellipticals from
\citep[][blue squares]{cho12}. \citet{peng08} red fractions are an
updated version of the results presented by \citet{peng06}, where red
fractions were corrected for the radial incompleteness that affects
mostly the higher-luminosity ellipticals in their
sample. Uncertainties have been taken from the original results of
\citet{peng06} since updated values are not given by
\citet{peng08}. Uncertainties for the red fractions measured by
\citet{cho12} are also not given in their paper.

The red fractions of Virgo galaxies present the same non-monotonic
behavior seen in \citet[][their Figure 8]{peng08} as function of $M_z$
instead of stellar mass. Red fractions increase from lower mass
galaxies up to a maximum around $\log M_{\star}\sim11$ beyond
which red fractions experience a turnover. Interestingly, the somewhat
lower red fractions for low-density ellipticals claimed by
\citet{cho12} are not seen; in the $10<\log M_{\star}<11$ range
the \citet{cho12} sample is indistinguishable from the Virgo
galaxies. 
That we cannot see lower red fractions for the \citet{cho12} sample
compared to the ACSVCS is probably based on, a) \citet{cho12} did not
use the red fractions corrected by aperture given by \citet{peng08},
instead using the uncorrected ones from \citet{peng06}, and b) the
comparison of red fractions was done as a function of $M_B$ instead of
stellar mass; for the same stellar mass, low-density ellipticals will
appear brighter in the optical as they are on average younger than
cluster ellipticals, so the claimed lower red fractions are only a
population effect on the galaxies luminosities and not an intrinsic
difference between the GCSs.

However, the galaxies presented in this paper appear to show
systematic lower fractions than the Virgo galaxies at similar stellar
mass, with the exception of IC4889.  IC 4889 has the shallowest
observations in our sample, thus the red fraction obtained for its GCS
would appear to be uncertain, accounting for its unusually high
value. It is important to note that, just like the values for the
total number of GCs, the $f_r$ are not corrected for unaccounted
clusters outside the FOV, based on the quick decline of the GC density
profiles (Sect. \ref{sec:radial}). If anything, wider field studies of
these isolated galaxies will reveal even lower red fractions. We
therefore conclude that red fractions in isolated ellipticals are
equal or possibly lower than the values for high-density ellipticals.

Recently, \citet{tonini13} introduced a model of GC formation based on
hierarchical merging and the mass-metallicity relation for
galaxies. An important feature of the model is the ability to make
predictions for galaxies with rich and poor accretion histories, where
poor-accretion galaxies end up with dominant red GC populations. If we
associate galaxies with poor accretion histories to present day
isolated ellipticals, our results are in strong disagreement with this
model and in line with the results of \citet{kartha14} for lenticular
galaxies. Additionally, it would be inconsistent with models
proposing an accreted origin from dwarf galaxies for the blue
population \citep[e.g.][]{cote98}.

A strong argument against the accreted origin of blue clusters is the
correlation between the peak color of the blue clusters and the
luminosity of the parent galaxy \citep[e.g.][]{larsen01,strader04},
which is not expected under an accretion scenario. Our finding of a
dominant blue peak despite the likely absence of a rich accretion
history, favours in-situ formation for GCs, where then the color
bimodality would be rather explained by a non-linear color-metallicity
relation in GCs \citep[e.g.][]{yoon06,yoon11}.

\subsection{The color peaks}

\begin{figure}
\includegraphics[width=0.48\textwidth]{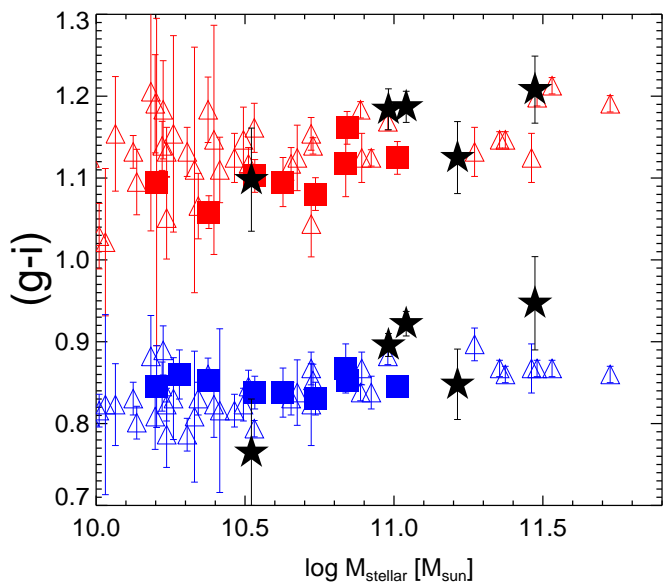}
\caption{Black stars indicate the red and blue peaks of the color
  distribution of GCs in our sample. Squares indicate galaxies from
  the \citet{cho12} sample, while open trinagles are the values from
  the ACSVCS \citep{peng08}. }
\label{fig:peaks}
\end{figure}

The mean color, or color peak, of the blue and red GC subpopulations
have both been found to be functions of the parent galaxy luminosity
\citep[e.g.][]{larsen01,strader04}, which suggest an intimate
connection between galaxy and GC enrichment history.

In order to compare again with \citet{cho12} and \citet{peng08}
samples, we transform $(g-z)$ colors from those studies into $(g-i)$
using the relation obtained by \citet{usher12} for 169 bonafide
cluster in NGC 4365,
\begin{equation}
(g - i)= (0.735 \pm 0.009) \times (g - z) + (0.147 \pm 0.012)\, \mathrm{mag} .
\end{equation}

Fig. \ref{fig:peaks} shows that on average both peak 
magnitudes in our sample are redder, by a small but significant 
amount, than the \citet{cho12} and \citet{peng08} peak magnitudes. 
This is in contrast to the results of \citet{cho12} who found that the 
peak magnitudes for their GCS were slightly bluer compared to the 
Virgo galaxies from \citet{peng08}, which also fits into the picture 
of isolated/field ellipticals having in average younger stellar 
populations. Given that our redder values go against this intuition, 
we believe our data may be affected by an error in its zeropoint 
photometry. We therefore restrict the analysis to an internal 
comparison between the galaxies in our sample.

Even though for the blue clusters, a slope may be more noticeable,
this is dominated by the most massive (NGC 3962), and least massive
galaxies (NGC 4240) of our sample, which are precisely the ones for
which color bimodality is not totally supported (see
Sect. \ref{sec:color}). Moreover, the mass--peak color relation for
blue clusters is expected to be even less pronounced for low-density
environments \citep{strader04}. We conclude that no relation with
galaxy mass is discernible for our sample, mostly limited by the
narrow range of masses studied.

\subsection{Ultra-compact dwarfs in NGC 3962?}
\label{sec:ucds}

The CMD of point sources around NGC 3962 shows a clear extension of
the GC sequence above the selected brighter limit
(Fig. \ref{fig:cmds}, top left panel, green triangles). Seventeen
point sources brighter than $i=21.24$ ($M_i=-11.5$ at the galaxy's
distance) and with $0.8<(g-i)_0<1.4$ can be seen. In this same range
of luminosity and colors only about three point sources are expected
based on the comparison field (the WHDF, Fig. \ref{fig:cmds}, bottom
right plot) and the Galactic model of \citet{robin03}. Stellar systems
of this bightness have been commonly labeled as ``ultra-compact
dwarfs'' \citep[UCDs, e.g.][]{drinkwater00}, due to the absence of
counterparts in the Local Group.

Even though some of the low luminosity UCD candidates can be just
bright GCs, given the uncertainty of 0.4 mag in the surface brightness
distance modulus measured by \citet{tonry01}, the correspondence
between the expected and measured peak of the GCLF in NGC 3962
(Fig. \ref{fig:lum_func}) shows that the true distance cannot be too off,
and hence most of these are genuinely brighter than the expected
brightest GCs.

UCDs have been found mostly in galaxy groups and clusters
\citep[e.g.][]{mieske08,brodie11}, with formation channels mostly
appealing to dynamical effects common in high-density environments
\citep[e.g.][]{hilker99}. If spectroscopically confirmed, the UCDs in
NGC 3962 would be the first around an isolated elliptical, confirming
that the formation of these objects is not restricted to high-density
environments, following the finding of one UCD around the field spiral
galaxy M 104 \citep{hau09}.

\section{Summary and conclusions}
\label{sec:conclusions}

In this paper we have studied the GCSs of five isolated elliptical
galaxies, NGC 2271, NGC 2865, NGC 3962, NGC 4240 and IC 4889, using
Gemini-S/GMOS $gi$ imaging. Our main results are:

\begin{itemize}
\item The GMM analysis indicates clear bimodality for three out of 5
  galaxies, while the two remaining show very short peak distances,
  making their bimodality less clear. In general, color bimodality is
  a common phenomenon down to the most isolated elliptical galaxies.

\item The specific frequency of isolated ellipticals, $S_N$, is close
  to 1.5 irrespectibly of the galaxy mass. In the range $10.5<\log
  M_{\star}<11.4$ a small correlation can be seen, in the sense that
  more massive isolated ellpiticals have smaller $S_N$, but for 
higher masses, larger $S_N$ are measured.

\item The conclusion of \citet{cho12} that galaxy mass is the main
  ruler of the properties of a GCS stems from the narrow range of
  masses used in their sample, biased towards low-luminosity
  galaxies. For galaxies with stellar masses $\log M_{\star}>11$,
  environment, and not mass, has the highest impact determining some
  of the global properties (e.g. specific frequencies) of the GCSs.

\item The red fractions show no significant difference with the ones
   seen in ellipticals in Virgo, making an accreted origin for the
  blue clusters doubtful. The slightly lower red fractions found by
  \citet{cho12} are a stellar population effect on the galaxies
  luminosities, and not intrinsic to the GCSs.
\end{itemize}

The last point assumes isolated ellipticals can be connected to poor
accretion history galaxies found in cosmological simulations. Even
though many isolated ellipticals show signs of \textit{recent}
mergers, to acquire knowledge of the entire accretion history, that
is, to establish whether present day isolated ellipticals have been
isolated throughout their lifetimes, is more difficult. As shown by
\citet{cooper13}, the outer surface brightness profiles of
high-density and low-density galaxies should be remarkably different
given their different accretion histories. The study of the outer
surface brightness profiles of isolated ellipticals as proxy of their
accretion histories will be the subject of forthcoming publications.


\begin{acknowledgements}

  We thank the anonymous referee for fast and valuable feedback on our
  manuscript. R.S. thanks Karla \'Alamo-Mart\'inez, Terry Bridges,
  Favio Faifer and especially Chiara Tonini for useful
  discussions. R.S. also thanks Andrei Tokovinin for his assistance
  during and after the SOAR/SAM observations. T.R. acknowledges
  financial support from FONDECYT project No. 1100620 and from the
  BASAL Centro de Astrof\'isica y Tecnolog\'ias Afines (CATA)
  PFB-06/2007. R.R.L. acknowledges financial support from FONDECYT
  project No. 3130403. Partly based on observations obtained at the
  Southern Astrophysical Research (SOAR) telescope, which is a joint
  project of the Minist\'{e}rio da Ci\^{e}ncia, Tecnologia, e
  Inova\c{c}\~{a}o (MCTI) da Rep\'{u}blica Federativa do Brasil, the
  U.S. National Optical Astronomy Observatory (NOAO), the University
  of North Carolina at Chapel Hill (UNC), and Michigan State
  University (MSU).
\end{acknowledgements}

\bibliographystyle{aa}
\bibliography{isolated}

\begin{thebibliography}{116}
\expandafter\ifx\csname natexlab\endcsname\relax\def\natexlab#1{#1}\fi

\bibitem[{{Alamo-Mart{\'{\i}}nez} {et~al.}(2012){Alamo-Mart{\'{\i}}nez},
  {West}, {Blakeslee}, {Gonz{\'a}lez-L{\'o}pezlira}, {Jord{\'a}n}, {Gregg},
  {C{\^o}t{\'e}}, {Drinkwater}, \& {van den Bergh}}]{alamo12}
{Alamo-Mart{\'{\i}}nez}, K.~A., {West}, M.~J., {Blakeslee}, J.~P., {et~al.}
  2012, \aap, 546, A15

\bibitem[{{Annibali} {et~al.}(2007){Annibali}, {Bressan}, {Rampazzo},
  {Zeilinger}, \& {Danese}}]{annibali07}
{Annibali}, F., {Bressan}, A., {Rampazzo}, R., {Zeilinger}, W.~W., \& {Danese},
  L. 2007, \aap, 463, 455

\bibitem[{{Ashman} {et~al.}(1994){Ashman}, {Bird}, \& {Zepf}}]{ashman94}
{Ashman}, K.~M., {Bird}, C.~M., \& {Zepf}, S.~E. 1994, \aj, 108, 2348

\bibitem[{{Ashman} \& {Zepf}(1992)}]{ashman92}
{Ashman}, K.~M. \& {Zepf}, S.~E. 1992, \apj, 384, 50

\bibitem[{{Bassino} {et~al.}(2006){Bassino}, {Faifer}, {Forte}, {Dirsch},
  {Richtler}, {Geisler}, \& {Schuberth}}]{bassino06}
{Bassino}, L.~P., {Faifer}, F.~R., {Forte}, J.~C., {et~al.} 2006, \aap, 451,
  789

\bibitem[{{Beasley} {et~al.}(2002){Beasley}, {Baugh}, {Forbes}, {Sharples}, \&
  {Frenk}}]{beasley02}
{Beasley}, M.~A., {Baugh}, C.~M., {Forbes}, D.~A., {Sharples}, R.~M., \&
  {Frenk}, C.~S. 2002, \mnras, 333, 383

\bibitem[{{Beers} {et~al.}(1990){Beers}, {Flynn}, \& {Gebhardt}}]{beers90}
{Beers}, T.~C., {Flynn}, K., \& {Gebhardt}, K. 1990, \aj, 100, 32

\bibitem[{{Bertin} \& {Arnouts}(1996)}]{bertin96}
{Bertin}, E. \& {Arnouts}, S. 1996, \aaps, 117, 393

\bibitem[{{B{\'{\i}}lek} {et~al.}(2014){B{\'{\i}}lek}, {Ebrov{\'a}},
  {Jungwiert}, {J{\'{\i}}lkov{\'a}}, \& {Barto{\v s}kov{\'a}}}]{bilek14}
{B{\'{\i}}lek}, M., {Ebrov{\'a}}, I., {Jungwiert}, B., {J{\'{\i}}lkov{\'a}},
  L., \& {Barto{\v s}kov{\'a}}, K. 2014, ArXiv e-prints

\bibitem[{{Blakeslee} {et~al.}(2012){Blakeslee}, {Cho}, {Peng}, {Ferrarese},
  {Jord{\'a}n}, \& {Martel}}]{blakeslee12}
{Blakeslee}, J.~P., {Cho}, H., {Peng}, E.~W., {et~al.} 2012, \apj, 746, 88

\bibitem[{{Blakeslee} {et~al.}(2009){Blakeslee}, {Jord{\'a}n}, {Mei},
  {C{\^o}t{\'e}}, {Ferrarese}, {Infante}, {Peng}, {Tonry}, \&
  {West}}]{blakeslee09}
{Blakeslee}, J.~P., {Jord{\'a}n}, A., {Mei}, S., {et~al.} 2009, \apj, 694, 556

\bibitem[{{Blakeslee} \& {Tonry}(1996)}]{blakeslee96}
{Blakeslee}, J.~P. \& {Tonry}, J.~L. 1996, \apjl, 465, L19

\bibitem[{{Brodie} {et~al.}(2011){Brodie}, {Romanowsky}, {Strader}, \&
  {Forbes}}]{brodie11}
{Brodie}, J.~P., {Romanowsky}, A.~J., {Strader}, J., \& {Forbes}, D.~A. 2011,
  \aj, 142, 199

\bibitem[{{Brodie} \& {Strader}(2006)}]{brodie06}
{Brodie}, J.~P. \& {Strader}, J. 2006, \araa, 44, 193

\bibitem[{{Brodie} {et~al.}(2012){Brodie}, {Usher}, {Conroy}, {Strader},
  {Arnold}, {Forbes}, \& {Romanowsky}}]{brodie12}
{Brodie}, J.~P., {Usher}, C., {Conroy}, C., {et~al.} 2012, \apjl, 759, L33

\bibitem[{{Bruzual} \& {Charlot}(2003)}]{bruzual03}
{Bruzual}, G. \& {Charlot}, S. 2003, \mnras, 344, 1000

\bibitem[{{Buson} {et~al.}(1993){Buson}, {Sadler}, {Zeilinger}, {Bertin},
  {Bertola}, {Danzinger}, {Dejonghe}, {Saglia}, \& {de Zeeuw}}]{buson93}
{Buson}, L.~M., {Sadler}, E.~M., {Zeilinger}, W.~W., {et~al.} 1993, \aap, 280,
  409

\bibitem[{{Caso} {et~al.}(2013){Caso}, {Richtler}, {Bassino}, {Salinas},
  {Lane}, \& {Romanowsky}}]{caso13}
{Caso}, J.~P., {Richtler}, T., {Bassino}, L.~P., {et~al.} 2013, \aap, 555, A56

\bibitem[{{Chies-Santos} {et~al.}(2012){Chies-Santos}, {Larsen}, {Cantiello},
  {Strader}, {Kuntschner}, {Wehner}, \& {Brodie}}]{chies-santos12}
{Chies-Santos}, A.~L., {Larsen}, S.~S., {Cantiello}, M., {et~al.} 2012, \aap,
  539, A54

\bibitem[{{Cho} {et~al.}(2012){Cho}, {Sharples}, {Blakeslee}, {Zepf}, {Kundu},
  {Kim}, \& {Yoon}}]{cho12}
{Cho}, J., {Sharples}, R.~M., {Blakeslee}, J.~P., {et~al.} 2012, \mnras, 422,
  3591

\bibitem[{{Clem} {et~al.}(2008){Clem}, {Vanden Berg}, \& {Stetson}}]{clem08}
{Clem}, J.~L., {Vanden Berg}, D.~A., \& {Stetson}, P.~B. 2008, \aj, 135, 682

\bibitem[{{Cole} {et~al.}(1994){Cole}, {Aragon-Salamanca}, {Frenk}, {Navarro},
  \& {Zepf}}]{cole94}
{Cole}, S., {Aragon-Salamanca}, A., {Frenk}, C.~S., {Navarro}, J.~F., \&
  {Zepf}, S.~E. 1994, \mnras, 271, 781

\bibitem[{{Collobert} {et~al.}(2006){Collobert}, {Sarzi}, {Davies},
  {Kuntschner}, \& {Colless}}]{collobert06}
{Collobert}, M., {Sarzi}, M., {Davies}, R.~L., {Kuntschner}, H., \& {Colless},
  M. 2006, \mnras, 370, 1213

\bibitem[{{Cooper} {et~al.}(2013){Cooper}, {D'Souza}, {Kauffmann}, {Wang},
  {Boylan-Kolchin}, {Guo}, {Frenk}, \& {White}}]{cooper13}
{Cooper}, A.~P., {D'Souza}, R., {Kauffmann}, G., {et~al.} 2013, \mnras, 434,
  3348

\bibitem[{{C\^ot\'e} {et~al.}(1998){C\^ot\'e}, {Marzke}, \& {West}}]{cote98}
{C\^ot\'e}, P., {Marzke}, R.~O., \& {West}, M.~J. 1998, \apj, 501, 554

\bibitem[{{De Lucia} {et~al.}(2006){De Lucia}, {Springel}, {White}, {Croton},
  \& {Kauffmann}}]{delucia06}
{De Lucia}, G., {Springel}, V., {White}, S.~D.~M., {Croton}, D., \&
  {Kauffmann}, G. 2006, \mnras, 366, 499

\bibitem[{{de Souza} {et~al.}(2004){de Souza}, {Gadotti}, \& {dos
  Anjos}}]{desouza04}
{de Souza}, R.~E., {Gadotti}, D.~A., \& {dos Anjos}, S. 2004, \apjs, 153, 411

\bibitem[{{Dirsch} {et~al.}(2003){Dirsch}, {Richtler}, {Geisler}, {Forte},
  {Bassino}, \& {Gieren}}]{dirsch03}
{Dirsch}, B., {Richtler}, T., {Geisler}, D., {et~al.} 2003, \aj, 125, 1908

\bibitem[{{Dressler}(1980)}]{dressler80}
{Dressler}, A. 1980, \apj, 236, 351

\bibitem[{{Drinkwater} {et~al.}(2000){Drinkwater}, {Jones}, {Gregg}, \&
  {Phillipps}}]{drinkwater00}
{Drinkwater}, M.~J., {Jones}, J.~B., {Gregg}, M.~D., \& {Phillipps}, S. 2000,
  \pasa, 17, 227

\bibitem[{{Durret} {et~al.}(2009){Durret}, {Slezak}, \& {Adami}}]{durret09}
{Durret}, F., {Slezak}, E., \& {Adami}, C. 2009, \aap, 506, 637

\bibitem[{{Elmegreen} {et~al.}(2012){Elmegreen}, {Malhotra}, \&
  {Rhoads}}]{elmegreen12}
{Elmegreen}, B.~G., {Malhotra}, S., \& {Rhoads}, J. 2012, \apj, 757, 9

\bibitem[{{Faifer} {et~al.}(2011){Faifer}, {Forte}, {Norris}, {Bridges},
  {Forbes}, {Zepf}, {Beasley}, {Gebhardt}, {Hanes}, \& {Sharples}}]{faifer11}
{Faifer}, F.~R., {Forte}, J.~C., {Norris}, M.~A., {et~al.} 2011, \mnras, 416,
  155

\bibitem[{{Fleming} {et~al.}(1995){Fleming}, {Harris}, {Pritchet}, \&
  {Hanes}}]{fleming95}
{Fleming}, D.~E.~B., {Harris}, W.~E., {Pritchet}, C.~J., \& {Hanes}, D.~A.
  1995, \aj, 109, 1044

\bibitem[{{Forbes} {et~al.}(1997){Forbes}, {Brodie}, \& {Grillmair}}]{forbes97}
{Forbes}, D.~A., {Brodie}, J.~P., \& {Grillmair}, C.~J. 1997, \aj, 113, 1652

\bibitem[{{Forbes} {et~al.}(2004){Forbes}, {Faifer}, {Forte}, {Bridges},
  {Beasley}, {Gebhardt}, {Hanes}, {Sharples}, \& {Zepf}}]{forbes04}
{Forbes}, D.~A., {Faifer}, F.~R., {Forte}, J.~C., {et~al.} 2004, \mnras, 355,
  608

\bibitem[{{Fort} {et~al.}(1986){Fort}, {Prieur}, {Carter}, {Meatheringham}, \&
  {Vigroux}}]{fort86}
{Fort}, B.~P., {Prieur}, J.-L., {Carter}, D., {Meatheringham}, S.~J., \&
  {Vigroux}, L. 1986, \apj, 306, 110

\bibitem[{{Fuse} {et~al.}(2012){Fuse}, {Marcum}, \& {Fanelli}}]{fuse12}
{Fuse}, C., {Marcum}, P., \& {Fanelli}, M. 2012, \aj, 144, 57

\bibitem[{{Gebhardt} \& {Kissler-Patig}(1999)}]{gebhardt99}
{Gebhardt}, K. \& {Kissler-Patig}, M. 1999, \aj, 118, 1526

\bibitem[{{Harris} {et~al.}(1991){Harris}, {Allwright}, {Pritchet}, \& {van den
  Bergh}}]{harris91}
{Harris}, W.~E., {Allwright}, J.~W.~B., {Pritchet}, C.~J., \& {van den Bergh},
  S. 1991, \apjs, 76, 115

\bibitem[{{Harris} {et~al.}(2013){Harris}, {Harris}, \& {Alessi}}]{harris13}
{Harris}, W.~E., {Harris}, G.~L.~H., \& {Alessi}, M. 2013, \apj, 772, 82

\bibitem[{{Harris} \& {van den Bergh}(1981)}]{harris81}
{Harris}, W.~E. \& {van den Bergh}, S. 1981, \aj, 86, 1627

\bibitem[{{Harris} {et~al.}(2006){Harris}, {Whitmore}, {Karakla}, {Oko{\'n}},
  {Baum}, {Hanes}, \& {Kavelaars}}]{harris06}
{Harris}, W.~E., {Whitmore}, B.~C., {Karakla}, D., {et~al.} 2006, \apj, 636, 90

\bibitem[{{Hau} {et~al.}(1999){Hau}, {Carter}, \& {Balcells}}]{hau99}
{Hau}, G.~K.~T., {Carter}, D., \& {Balcells}, M. 1999, \mnras, 306, 437

\bibitem[{{Hau} \& {Forbes}(2006)}]{hau06}
{Hau}, G.~K.~T. \& {Forbes}, D.~A. 2006, \mnras, 371, 633

\bibitem[{{Hau} {et~al.}(2009){Hau}, {Spitler}, {Forbes}, {Proctor}, {Strader},
  {Mendel}, {Brodie}, \& {Harris}}]{hau09}
{Hau}, G.~K.~T., {Spitler}, L.~R., {Forbes}, D.~A., {et~al.} 2009, \mnras, 394,
  L97

\bibitem[{{Hernquist} \& {Quinn}(1987)}]{hernquist87}
{Hernquist}, L. \& {Quinn}, P.~J. 1987, \apj, 312, 1

\bibitem[{{Hilker} {et~al.}(1999){Hilker}, {Infante}, \& {Richtler}}]{hilker99}
{Hilker}, M., {Infante}, L., \& {Richtler}, T. 1999, \aaps, 138, 55

\bibitem[{{Ibata} {et~al.}(1994){Ibata}, {Gilmore}, \& {Irwin}}]{ibata94}
{Ibata}, R.~A., {Gilmore}, G., \& {Irwin}, M.~J. 1994, \nat, 370, 194

\bibitem[{{Izenmann}(1991)}]{izenmann91}
{Izenmann}, A.~J. 1991, Am. Stat. Assoc., 86, 205

\bibitem[{{Jarrett} {et~al.}(2003){Jarrett}, {Chester}, {Cutri}, {Schneider},
  \& {Huchra}}]{jarrett03}
{Jarrett}, T.~H., {Chester}, T., {Cutri}, R., {Schneider}, S.~E., \& {Huchra},
  J.~P. 2003, \aj, 125, 525

\bibitem[{{Kartha} {et~al.}(2014){Kartha}, {Forbes}, {Spitler}, {Romanowsky},
  {Arnold}, \& {Brodie}}]{kartha14}
{Kartha}, S.~S., {Forbes}, D.~A., {Spitler}, L.~R., {et~al.} 2014, \mnras, 437,
  273

\bibitem[{{Kundu} \& {Whitmore}(2001)}]{kundu01}
{Kundu}, A. \& {Whitmore}, B.~C. 2001, \aj, 121, 2950

\bibitem[{{Kuntschner} {et~al.}(2002){Kuntschner}, {Smith}, {Colless},
  {Davies}, {Kaldare}, \& {Vazdekis}}]{kuntschner02}
{Kuntschner}, H., {Smith}, R.~J., {Colless}, M., {et~al.} 2002, \mnras, 337,
  172

\bibitem[{{La Barbera} {et~al.}(2012){La Barbera}, {Ferreras}, {de Carvalho},
  {Bruzual}, {Charlot}, {Pasquali}, \& {Merlin}}]{labarbera12}
{La Barbera}, F., {Ferreras}, I., {de Carvalho}, R.~R., {et~al.} 2012, \mnras,
  426, 2300

\bibitem[{{Lane} {et~al.}(2011){Lane}, {Kiss}, {Lewis}, {Ibata}, {Siebert},
  {Bedding}, {Sz{\'e}kely}, \& {Szab{\'o}}}]{lane11}
{Lane}, R.~R., {Kiss}, L.~L., {Lewis}, G.~F., {et~al.} 2011, \aap, 530, A31

\bibitem[{{Lane} {et~al.}(2013){Lane}, {Salinas}, \& {Richtler}}]{lane13}
{Lane}, R.~R., {Salinas}, R., \& {Richtler}, T. 2013, \aap, 549, A148

\bibitem[{{Larsen} {et~al.}(2001){Larsen}, {Brodie}, {Huchra}, {Forbes}, \&
  {Grillmair}}]{larsen01}
{Larsen}, S.~S., {Brodie}, J.~P., {Huchra}, J.~P., {Forbes}, D.~A., \&
  {Grillmair}, C.~J. 2001, \aj, 121, 2974

\bibitem[{{Laurikainen} {et~al.}(2011){Laurikainen}, {Salo}, {Buta}, \&
  {Knapen}}]{laurikainen11}
{Laurikainen}, E., {Salo}, H., {Buta}, R., \& {Knapen}, J.~H. 2011, \mnras,
  418, 1452

\bibitem[{{Liu} {et~al.}(2011){Liu}, {Peng}, {Jord{\'a}n}, {Ferrarese},
  {Blakeslee}, {C{\^o}t{\'e}}, \& {Mei}}]{liu11}
{Liu}, C., {Peng}, E.~W., {Jord{\'a}n}, A., {et~al.} 2011, \apj, 728, 116

\bibitem[{{Madore} {et~al.}(2004){Madore}, {Freedman}, \& {Bothun}}]{madore04}
{Madore}, B.~F., {Freedman}, W.~L., \& {Bothun}, G.~D. 2004, \apj, 607, 810

\bibitem[{{Malin} \& {Carter}(1983)}]{malin83}
{Malin}, D.~F. \& {Carter}, D. 1983, \apj, 274, 534

\bibitem[{{Markwardt}(2009)}]{markwardt09}
{Markwardt}, C.~B. 2009, in Astronomical Society of the Pacific Conference
  Series, Vol. 411, Astronomical Data Analysis Software and Systems XVIII, ed.
  D.~A. {Bohlender}, D.~{Durand}, \& P.~{Dowler}, 251

\bibitem[{{McLaughlin} {et~al.}(1994){McLaughlin}, {Harris}, \&
  {Hanes}}]{mclaughlin94}
{McLaughlin}, D.~E., {Harris}, W.~E., \& {Hanes}, D.~A. 1994, \apj, 422, 486

\bibitem[{{Metcalfe} {et~al.}(2001){Metcalfe}, {Shanks}, {Campos}, {McCracken},
  \& {Fong}}]{metcalfe01}
{Metcalfe}, N., {Shanks}, T., {Campos}, A., {McCracken}, H.~J., \& {Fong}, R.
  2001, \mnras, 323, 795

\bibitem[{{Michard} \& {Prugniel}(2004)}]{michard04}
{Michard}, R. \& {Prugniel}, P. 2004, \aap, 423, 833

\bibitem[{{Mieske} {et~al.}(2008){Mieske}, {Hilker}, {Jord{\'a}n}, {Infante},
  {Kissler-Patig}, {Rejkuba}, {Richtler}, {C{\^o}t{\'e}}, {Baumgardt}, {West},
  {Ferrarese}, \& {Peng}}]{mieske08}
{Mieske}, S., {Hilker}, M., {Jord{\'a}n}, A., {et~al.} 2008, \aap, 487, 921

\bibitem[{{Muratov} \& {Gnedin}(2010)}]{muratov10}
{Muratov}, A.~L. \& {Gnedin}, O.~Y. 2010, \apj, 718, 1266

\bibitem[{{Niemi} {et~al.}(2010){Niemi}, {Hein{\"a}m{\"a}ki}, {Nurmi}, \&
  {Saar}}]{niemi10}
{Niemi}, S., {Hein{\"a}m{\"a}ki}, P., {Nurmi}, P., \& {Saar}, E. 2010, \mnras,
  471

\bibitem[{{Nigoche-Netro} {et~al.}(2007){Nigoche-Netro}, {Moles},
  {Ruelas-Mayorga}, {Franco-Balderas}, \& {Kj{\o}rgaard}}]{nigoche07}
{Nigoche-Netro}, A., {Moles}, M., {Ruelas-Mayorga}, A., {Franco-Balderas}, A.,
  \& {Kj{\o}rgaard}, P. 2007, \aap, 472, 773

\bibitem[{{Peng} {et~al.}(2006){Peng}, {Jord{\'a}n}, {C{\^o}t{\'e}},
  {Blakeslee}, {Ferrarese}, {Mei}, {West}, {Merritt}, {Milosavljevi{\'c}}, \&
  {Tonry}}]{peng06}
{Peng}, E.~W., {Jord{\'a}n}, A., {C{\^o}t{\'e}}, P., {et~al.} 2006, \apj, 639,
  95

\bibitem[{{Peng} {et~al.}(2008){Peng}, {Jord{\'a}n}, {C{\^o}t{\'e}},
  {Takamiya}, {West}, {Blakeslee}, {Chen}, {Ferrarese}, {Mei}, {Tonry}, \&
  {West}}]{peng08}
{Peng}, E.~W., {Jord{\'a}n}, A., {C{\^o}t{\'e}}, P., {et~al.} 2008, \apj, 681,
  197

\bibitem[{{Reda} {et~al.}(2004){Reda}, {Forbes}, {Beasley}, {O'Sullivan}, \&
  {Goudfrooij}}]{reda04}
{Reda}, F.~M., {Forbes}, D.~A., {Beasley}, M.~A., {O'Sullivan}, E.~J., \&
  {Goudfrooij}, P. 2004, \mnras, 354, 851

\bibitem[{{Reda} {et~al.}(2007){Reda}, {Proctor}, {Forbes}, {Hau}, \&
  {Larsen}}]{reda07}
{Reda}, F.~M., {Proctor}, R.~N., {Forbes}, D.~A., {Hau}, G.~K.~T., \& {Larsen},
  S.~S. 2007, \mnras, 377, 1772

\bibitem[{{Rejkuba}(2012)}]{rejkuba12}
{Rejkuba}, M. 2012, \apss, 341, 195

\bibitem[{{Rhode} \& {Zepf}(2001)}]{rhode01}
{Rhode}, K.~L. \& {Zepf}, S.~E. 2001, \aj, 121, 210

\bibitem[{{Rhode} {et~al.}(2005){Rhode}, {Zepf}, \& {Santos}}]{rhode05}
{Rhode}, K.~L., {Zepf}, S.~E., \& {Santos}, M.~R. 2005, \apjl, 630, L21

\bibitem[{{Richtler}(2003)}]{richtler03}
{Richtler}, T. 2003, in Lecture Notes in Physics, Berlin Springer Verlag, Vol.
  635, Stellar Candles for the Extragalactic Distance Scale, ed. D.~{Alloin} \&
  W.~{Gieren}, 281--305

\bibitem[{{Richtler}(2006)}]{richtler06}
{Richtler}, T. 2006, Bulletin of the Astronomical Society of India, 34, 83

\bibitem[{{Richtler}(2013)}]{richtler13}
{Richtler}, T. 2013, in Astronomical Society of the Pacific Conference Series,
  Vol. 470, 370 Years of Astronomy in Utrecht, ed. G.~{Pugliese}, A.~{de
  Koter}, \& M.~{Wijburg}, 327

\bibitem[{{Richtler} {et~al.}(2012){Richtler}, {Bassino}, {Dirsch}, \&
  {Kumar}}]{richtler12}
{Richtler}, T., {Bassino}, L.~P., {Dirsch}, B., \& {Kumar}, B. 2012, \aap, 543,
  A131

\bibitem[{{Richtler} {et~al.}(2015){Richtler}, {Salinas}, {Lane}, {Hilker}, \&
  {Schirmer}}]{richtler15}
{Richtler}, T., {Salinas}, R., {Lane}, R.~R., {Hilker}, M., \& {Schirmer}, M.
  2015, \aap, 574, A21

\bibitem[{{Richtler} {et~al.}(2011){Richtler}, {Salinas}, {Misgeld}, {Hilker},
  {Hau}, {Romanowsky}, {Schuberth}, \& {Spolaor}}]{richtler11}
{Richtler}, T., {Salinas}, R., {Misgeld}, I., {et~al.} 2011, \aap, 531, A119

\bibitem[{{Robin} {et~al.}(2003){Robin}, {Reyl{\'e}}, {Derri{\`e}re}, \&
  {Picaud}}]{robin03}
{Robin}, A.~C., {Reyl{\'e}}, C., {Derri{\`e}re}, S., \& {Picaud}, S. 2003,
  \aap, 409, 523

\bibitem[{{Salinas} {et~al.}(2012){Salinas}, {Richtler}, {Bassino},
  {Romanowsky}, \& {Schuberth}}]{salinas12}
{Salinas}, R., {Richtler}, T., {Bassino}, L.~P., {Romanowsky}, A.~J., \&
  {Schuberth}, Y. 2012, \aap, 538, A87

\bibitem[{{S{\'a}nchez-Bl{\'a}zquez} {et~al.}(2007){S{\'a}nchez-Bl{\'a}zquez},
  {Forbes}, {Strader}, {Brodie}, \& {Proctor}}]{sanchez07}
{S{\'a}nchez-Bl{\'a}zquez}, P., {Forbes}, D.~A., {Strader}, J., {Brodie}, J.,
  \& {Proctor}, R. 2007, \mnras, 377, 759

\bibitem[{{Sarzi} {et~al.}(2006){Sarzi}, {Falc{\'o}n-Barroso}, {Davies},
  {Bacon}, {Bureau}, {Cappellari}, {de Zeeuw}, {Emsellem}, {Fathi},
  {Krajnovi{\'c}}, {Kuntschner}, {McDermid}, \& {Peletier}}]{sarzi06}
{Sarzi}, M., {Falc{\'o}n-Barroso}, J., {Davies}, R.~L., {et~al.} 2006, \mnras,
  366, 1151

\bibitem[{{Schlafly} \& {Finkbeiner}(2011)}]{schlafly11}
{Schlafly}, E.~F. \& {Finkbeiner}, D.~P. 2011, \apj, 737, 103

\bibitem[{{Schlegel} {et~al.}(1998){Schlegel}, {Finkbeiner}, \&
  {Davis}}]{schlegel98}
{Schlegel}, D.~J., {Finkbeiner}, D.~P., \& {Davis}, M. 1998, \apj, 500, 525

\bibitem[{{Schuberth} {et~al.}(2012){Schuberth}, {Richtler}, {Hilker},
  {Salinas}, {Dirsch}, \& {Larsen}}]{schuberth12}
{Schuberth}, Y., {Richtler}, T., {Hilker}, M., {et~al.} 2012, \aap, 544, A115

\bibitem[{{Serra} \& {Oosterloo}(2010)}]{serra10}
{Serra}, P. \& {Oosterloo}, T.~A. 2010, \mnras, 401, L29

\bibitem[{{Sikkema} {et~al.}(2006){Sikkema}, {Peletier}, {Carter}, {Valentijn},
  \& {Balcells}}]{sikkema06}
{Sikkema}, G., {Peletier}, R.~F., {Carter}, D., {Valentijn}, E.~A., \&
  {Balcells}, M. 2006, \aap, 458, 53

\bibitem[{{Smith} {et~al.}(2002){Smith}, {Tucker}, {Kent}, {Richmond},
  {Fukugita}, {Ichikawa}, {Ichikawa}, {Jorgensen}, {Uomoto}, {Gunn}, {Hamabe},
  {Watanabe}, {Tolea}, {Henden}, {Annis}, {Pier}, {McKay}, {Brinkmann}, {Chen},
  {Holtzman}, {Shimasaku}, \& {York}}]{smith02}
{Smith}, J.~A., {Tucker}, D.~L., {Kent}, S., {et~al.} 2002, \aj, 123, 2121

\bibitem[{{Smith} {et~al.}(2004){Smith}, {Mart{\'{\i}}nez}, \&
  {Graham}}]{smith04}
{Smith}, R.~M., {Mart{\'{\i}}nez}, V.~J., \& {Graham}, M.~J. 2004, \apj, 617,
  1017

\bibitem[{{Spitler} {et~al.}(2008){Spitler}, {Forbes}, {Strader}, {Brodie}, \&
  {Gallagher}}]{spitler08}
{Spitler}, L.~R., {Forbes}, D.~A., {Strader}, J., {Brodie}, J.~P., \&
  {Gallagher}, J.~S. 2008, \mnras, 385, 361

\bibitem[{{Stetson}(1987)}]{stetson87}
{Stetson}, P.~B. 1987, \pasp, 99, 191

\bibitem[{{Stetson}(1993)}]{stetson93}
{Stetson}, P.~B. 1993, in IAU Colloq. 136: Stellar Photometry - Current
  Techniques and Future Developments, ed. {C.~J.~Butler \& I.~Elliott}, 291

\bibitem[{{Stocke} {et~al.}(2004){Stocke}, {Keeney}, {Lewis}, {Epps}, \&
  {Schild}}]{stocke04}
{Stocke}, J.~T., {Keeney}, B.~A., {Lewis}, A.~D., {Epps}, H.~W., \& {Schild},
  R.~E. 2004, \aj, 127, 1336

\bibitem[{{Strader} {et~al.}(2007){Strader}, {Beasley}, \&
  {Brodie}}]{strader07}
{Strader}, J., {Beasley}, M.~A., \& {Brodie}, J.~P. 2007, \aj, 133, 2015

\bibitem[{{Strader} {et~al.}(2004){Strader}, {Brodie}, \& {Forbes}}]{strader04}
{Strader}, J., {Brodie}, J.~P., \& {Forbes}, D.~A. 2004, \aj, 127, 3431

\bibitem[{{Strader} {et~al.}(2006){Strader}, {Brodie}, {Spitler}, \&
  {Beasley}}]{strader06}
{Strader}, J., {Brodie}, J.~P., {Spitler}, L., \& {Beasley}, M.~A. 2006, \aj,
  132, 2333

\bibitem[{{Tal} {et~al.}(2009){Tal}, {van Dokkum}, {Nelan}, \&
  {Bezanson}}]{tal09}
{Tal}, T., {van Dokkum}, P.~G., {Nelan}, J., \& {Bezanson}, R. 2009, \aj, 138,
  1417

\bibitem[{{Tokovinin} {et~al.}(2010){Tokovinin}, {Cantarutti}, {Tighe},
  {Schurter}, {van der Bliek}, {Martinez}, \& {Mondaca}}]{tokovinin10}
{Tokovinin}, A., {Cantarutti}, R., {Tighe}, R., {et~al.} 2010, \pasp, 122, 1483

\bibitem[{{Tokovinin} {et~al.}(2012){Tokovinin}, {Tighe}, {Schurter},
  {Cantarutti}, {van der Bliek}, {Martinez}, {Mondaca}, \&
  {Heathcote}}]{tokovinin12}
{Tokovinin}, A., {Tighe}, R., {Schurter}, P., {et~al.} 2012, in Society of
  Photo-Optical Instrumentation Engineers (SPIE) Conferenc e Series, Vol. 8447,
  Society of Photo-Optical Instrumentation Engineers (SPIE) Conferenc e Series

\bibitem[{{Tonini}(2013)}]{tonini13}
{Tonini}, C. 2013, \apj, 762, 39

\bibitem[{{Tonry} {et~al.}(2001){Tonry}, {Dressler}, {Blakeslee}, {Ajhar},
  {Fletcher}, {Luppino}, {Metzger}, \& {Moore}}]{tonry01}
{Tonry}, J.~L., {Dressler}, A., {Blakeslee}, J.~P., {et~al.} 2001, \apj, 546,
  681

\bibitem[{{Trancho} {et~al.}(2014){Trancho}, {Miller}, {Schweizer}, {Burdett},
  \& {Palamara}}]{trancho14}
{Trancho}, G., {Miller}, B.~W., {Schweizer}, F., {Burdett}, D.~P., \&
  {Palamara}, D. 2014, \apj, 790, 122

\bibitem[{{Tully}(1988)}]{tully88}
{Tully}, R.~B. 1988, {Nearby galaxies catalog}

\bibitem[{{Urrutia-Viscarra} {et~al.}(2014){Urrutia-Viscarra}, {Arnaboldi},
  {Mendes de Oliveira}, {Gerhard}, {Torres-Flores}, {Carrasco}, \& {de
  Mello}}]{urrutia14}
{Urrutia-Viscarra}, F., {Arnaboldi}, M., {Mendes de Oliveira}, C., {et~al.}
  2014, \aap, 569, A97

\bibitem[{{Usher} {et~al.}(2012){Usher}, {Forbes}, {Brodie}, {Foster},
  {Spitler}, {Arnold}, {Romanowsky}, {Strader}, \& {Pota}}]{usher12}
{Usher}, C., {Forbes}, D.~A., {Brodie}, J.~P., {et~al.} 2012, \mnras, 426, 1475

\bibitem[{{van Dokkum} {et~al.}(2010){van Dokkum}, {Whitaker}, {Brammer},
  {Franx}, {Kriek}, {Labb{\'e}}, {Marchesini}, {Quadri}, {Bezanson},
  {Illingworth}, {Muzzin}, {Rudnick}, {Tal}, \& {Wake}}]{vandokkum10}
{van Dokkum}, P.~G., {Whitaker}, K.~E., {Brammer}, G., {et~al.} 2010, \apj,
  709, 1018

\bibitem[{{Villegas} {et~al.}(2010){Villegas}, {Jord{\'a}n}, {Peng},
  {Blakeslee}, {C{\^o}t{\'e}}, {Ferrarese}, {Kissler-Patig}, {Mei}, {Infante},
  {Tonry}, \& {West}}]{villegas10}
{Villegas}, D., {Jord{\'a}n}, A., {Peng}, E.~W., {et~al.} 2010, \apj, 717, 603

\bibitem[{{Wehner} {et~al.}(2008){Wehner}, {Harris}, {Whitmore}, {Rothberg}, \&
  {Woodley}}]{wehner08}
{Wehner}, E.~M.~H., {Harris}, W.~E., {Whitmore}, B.~C., {Rothberg}, B., \&
  {Woodley}, K.~A. 2008, \apj, 681, 1233

\bibitem[{{Yoon} {et~al.}(2011){Yoon}, {Lee}, {Blakeslee}, {Peng}, {Sohn},
  {Cho}, {Kim}, {Chung}, {Kim}, \& {Lee}}]{yoon11}
{Yoon}, S.-J., {Lee}, S.-Y., {Blakeslee}, J.~P., {et~al.} 2011, \apj, 743, 150

\bibitem[{{Yoon} {et~al.}(2006){Yoon}, {Yi}, \& {Lee}}]{yoon06}
{Yoon}, S.-J., {Yi}, S.~K., \& {Lee}, Y.-W. 2006, Science, 311, 1129

\bibitem[{{Zepf} \& {Ashman}(1993)}]{zepf93}
{Zepf}, S.~E. \& {Ashman}, K.~M. 1993, \mnras, 264, 611

\end{thebibliography}

\begin{appendix}
\section{The color maps}
\label{sec:colormaps}
$g-i$ color maps are given for all five galaxies. Each panel has a
different angular scale, appropriate to indicate relevant color
features in each galaxy. For all figures North is up, East, to the
left.

\begin{figure}
\includegraphics[width=0.49\textwidth]{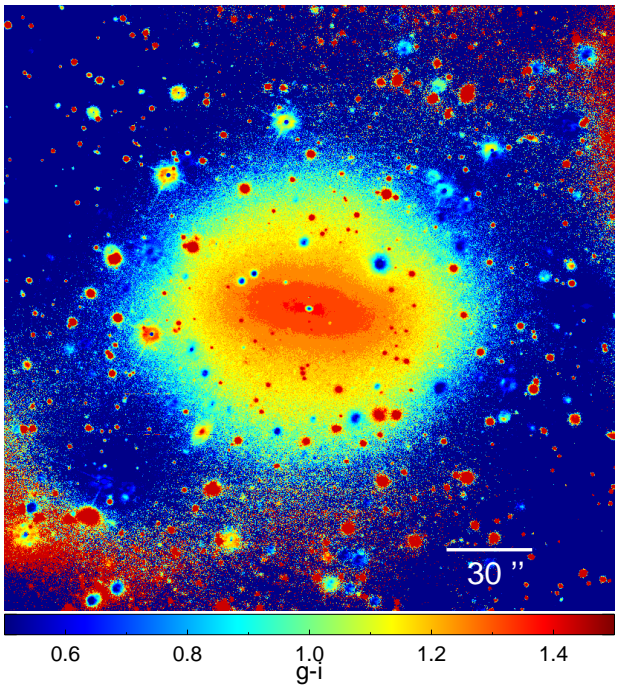}
\caption{NGC 2271 color map. }
\label{fig:colorn2271}
\end{figure}

\begin{figure*}
\includegraphics[width=\textwidth]{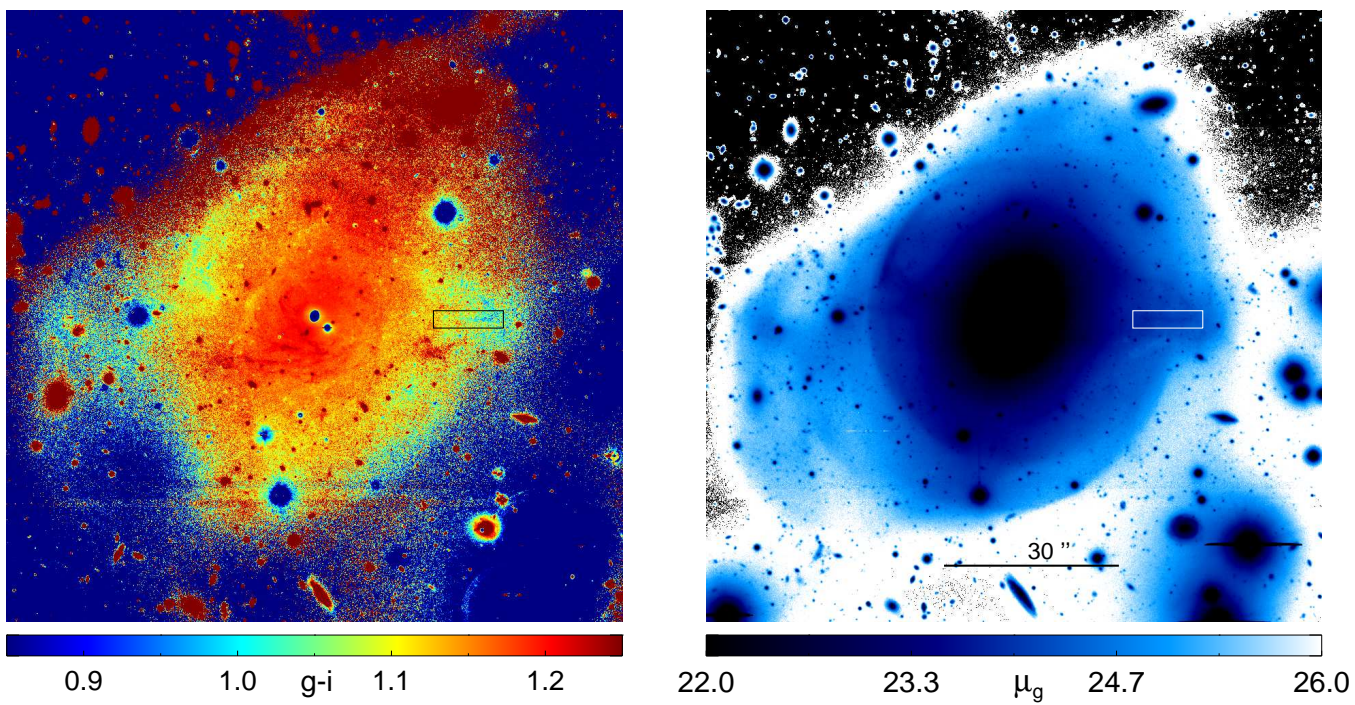}
\caption{{\it Left panel}:$g-i$ color map of NGC 2865. {\it Right
    panel}: $g$ surface brightness map of NGC 2865. Both figures share
  the same size and scale. The black/white box indicates the radial 
feature discussed in the text.}
\label{fig:colorn2865}
\end{figure*}

\begin{figure*}
\includegraphics[width=\textwidth]{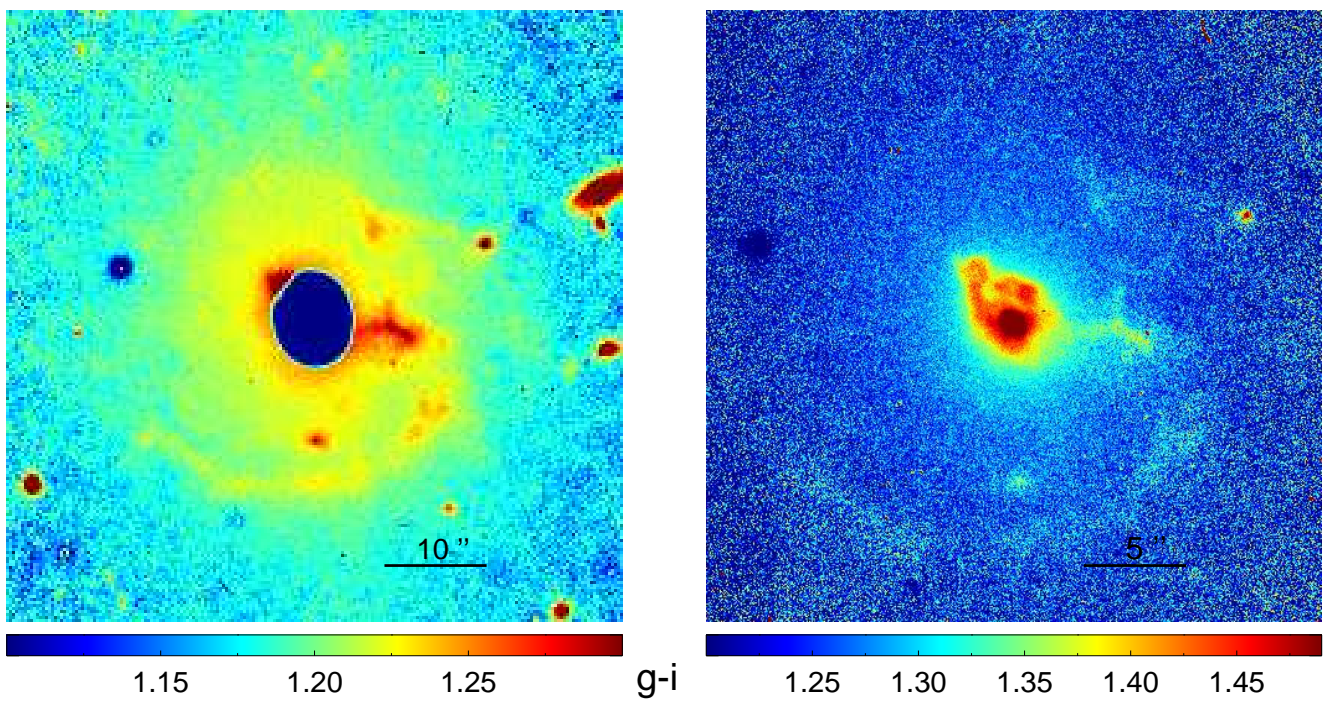}
\caption{$g-i$ color map for the central parts of NGC 3962. {\it Left
    panel}: Gemini/GMOS map, $1\arcmin$ aside. The central blue area
  is due to detector saturation. {\it Right panel}: SOAR/SAM color
  map, 30\arcsec aside. Albeit with lower $S/N$, it shows details
  unseen in the Gemini image due to saturation.}
\label{fig:colorn3962}
\end{figure*}

\begin{figure}
\includegraphics[width=0.49\textwidth]{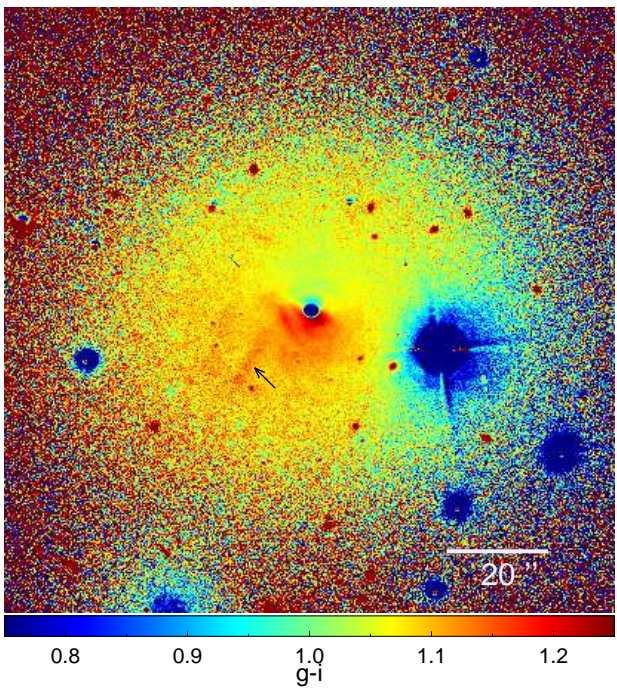}
\caption{NGC 4240 color map. }
\label{fig:colorn4240}
\end{figure}

\begin{figure}
\includegraphics[width=0.49\textwidth]{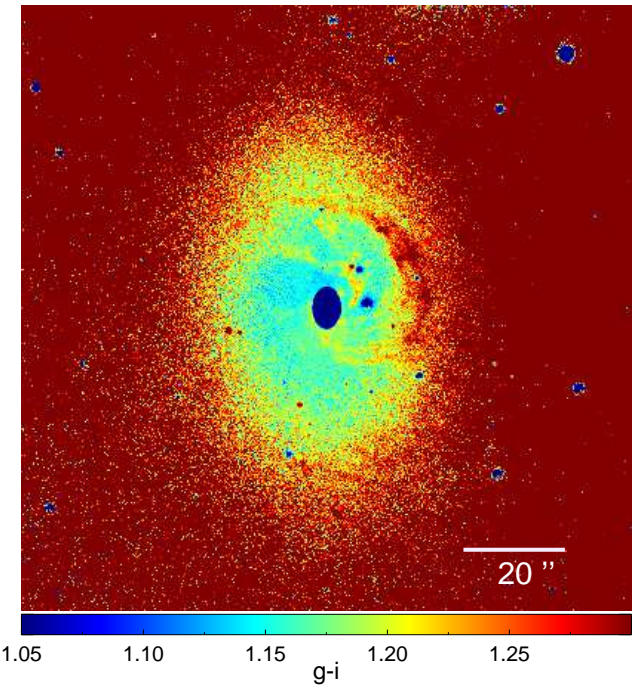}
\caption{IC 4889 color map. }
\label{fig:coloric4889}
\end{figure}

\newpage

\section{The surface brightness profiles}
\label{sec:surface}
$g$ and $i$ surface brightness profiles are given for each galaxy as
measured with iraf/ellipse. The indicated radii are the mean geometric
radii of each fitted ellipse. Radii are given in arc seconds, while
surface brightness, in mag/arcsec$^2$.

\begin{table}
\tiny
\caption{NGC 2271 surface photometry}
\label{table:n2271sb}
 \centering
  \begin{tabular}{@{}lcccc@{}}
    \hline\hline
$R$  &   $g$  & $g_{\rm err}$& $i$ & $i_{\rm err}$\\
\hline
2.36 & 18.630 & 0.000 & 17.352 & 0.000 \\
2.58 & 18.719 & 0.000 & 17.441 & 0.000 \\
2.82 & 18.810 & 0.000 & 17.531 & 0.000 \\
3.08 & 18.902 & 0.000 & 17.621 & 0.000 \\
3.36 & 18.995 & 0.000 & 17.715 & 0.001 \\
3.66 & 19.091 & 0.000 & 17.813 & 0.001 \\
4.00 & 19.189 & 0.000 & 17.912 & 0.001 \\
4.35 & 19.287 & 0.000 & 18.013 & 0.001 \\
4.73 & 19.387 & 0.000 & 18.114 & 0.001 \\
5.16 & 19.495 & 0.000 & 18.223 & 0.001 \\
5.63 & 19.612 & 0.000 & 18.341 & 0.001 \\
6.16 & 19.738 & 0.000 & 18.469 & 0.001 \\
6.74 & 19.869 & 0.000 & 18.597 & 0.001 \\
7.35 & 19.997 & 0.000 & 18.731 & 0.001 \\
8.03 & 20.131 & 0.000 & 18.869 & 0.001 \\
8.79 & 20.273 & 0.000 & 19.012 & 0.002 \\
9.58 & 20.413 & 0.000 & 19.156 & 0.002 \\
10.48 & 20.569 & 0.001 & 19.315 & 0.002 \\
11.46 & 20.733 & 0.001 & 19.482 & 0.003 \\
12.55 & 20.911 & 0.001 & 19.664 & 0.003 \\
13.77 & 21.103 & 0.001 & 19.861 & 0.004 \\
15.08 & 21.295 & 0.001 & 20.054 & 0.004 \\
16.66 & 21.522 & 0.001 & 20.288 & 0.005 \\
18.44 & 21.747 & 0.002 & 20.521 & 0.007 \\
20.44 & 21.976 & 0.002 & 20.760 & 0.008 \\
22.93 & 22.226 & 0.002 & 21.013 & 0.011 \\
25.75 & 22.483 & 0.003 & 21.288 & 0.014 \\
28.98 & 22.732 & 0.004 & 21.550 & 0.017 \\
32.47 & 22.969 & 0.005 & 21.795 & 0.022 \\
36.39 & 23.211 & 0.006 & 22.058 & 0.028 \\
40.43 & 23.435 & 0.007 & 22.305 & 0.035 \\
44.47 & 23.649 & 0.009 & 22.520 & 0.042 \\
48.92 & 23.868 & 0.011 & 22.747 & 0.052 \\
53.81 & 24.080 & 0.013 & 22.988 & 0.064 \\
59.19 & 24.316 & 0.016 & 23.223 & 0.079 \\
65.11 & 24.572 & 0.021 & 23.500 & 0.101 \\
71.63 & 24.781 & 0.025 & 23.663 & 0.117 \\
78.79 & 24.898 & 0.028 & 23.885 & 0.142 \\
86.67 & 25.201 & 0.036 & 24.123 & 0.174 \\
95.33 & 25.349 & 0.042 & 24.288 & 0.200 \\
104.87 & 25.547 & 0.050 & 24.375 & 0.215 \\
115.35 & 25.558 & 0.050 & 24.390 & 0.218 \\
\hline
\end{tabular}
\end{table}

\begin{table}
\tiny
\caption{NGC 2865 surface photometry}
\label{table:n2865sb}
 \centering
  \begin{tabular}{@{}lcccc@{}}
    \hline\hline
$R$  &   $g$  & $g_{\rm err}$& $i$ & $i_{\rm err}$\\
\hline
3.65 & 18.557 & 0.000 & 17.547 & 0.000 \\
3.99 & 18.681 & 0.000 & 17.665 & 0.000 \\
4.37 & 18.808 & 0.000 & 17.784 & 0.000 \\
4.77 & 18.938 & 0.000 & 17.906 & 0.000 \\
5.22 & 19.073 & 0.000 & 18.038 & 0.000 \\
5.72 & 19.210 & 0.000 & 18.172 & 0.000 \\
6.20 & 19.340 & 0.000 & 18.305 & 0.000 \\
6.77 & 19.476 & 0.000 & 18.444 & 0.000 \\
7.40 & 19.621 & 0.000 & 18.581 & 0.000 \\
8.09 & 19.764 & 0.000 & 18.727 & 0.000 \\
8.85 & 19.905 & 0.000 & 18.871 & 0.000 \\
9.65 & 20.039 & 0.001 & 19.001 & 0.000 \\
10.53 & 20.168 & 0.001 & 19.136 & 0.000 \\
11.56 & 20.308 & 0.001 & 19.274 & 0.001 \\
12.68 & 20.458 & 0.001 & 19.424 & 0.001 \\
14.02 & 20.633 & 0.001 & 19.594 & 0.001 \\
15.52 & 20.819 & 0.001 & 19.785 & 0.001 \\
17.17 & 21.011 & 0.001 & 19.987 & 0.001 \\
18.95 & 21.212 & 0.002 & 20.189 & 0.001 \\
21.00 & 21.428 & 0.002 & 20.410 & 0.001 \\
23.19 & 21.648 & 0.002 & 20.632 & 0.002 \\
25.91 & 21.888 & 0.003 & 20.874 & 0.002 \\
28.99 & 22.126 & 0.004 & 21.121 & 0.003 \\
31.63 & 22.310 & 0.004 & 21.310 & 0.003 \\
34.64 & 22.519 & 0.005 & 21.522 & 0.004 \\
38.28 & 22.767 & 0.006 & 21.762 & 0.005 \\
42.17 & 22.987 & 0.008 & 21.983 & 0.006 \\
46.38 & 23.164 & 0.009 & 22.175 & 0.007 \\
51.02 & 23.363 & 0.011 & 22.378 & 0.009 \\
56.12 & 23.554 & 0.013 & 22.583 & 0.011 \\
61.73 & 23.758 & 0.016 & 22.793 & 0.013 \\
67.91 & 24.007 & 0.020 & 23.026 & 0.016 \\
74.70 & 24.300 & 0.026 & 23.333 & 0.021 \\
82.17 & 24.616 & 0.034 & 23.671 & 0.029 \\
90.39 & 25.054 & 0.051 & 24.099 & 0.043 \\
99.42 & 25.436 & 0.072 & 24.586 & 0.067 \\
109.37 & 25.627 & 0.086 & 24.885 & 0.087 \\
120.30 & 25.880 & 0.107 & 25.081 & 0.104 \\
132.33 & 26.171 & 0.140 & 25.564 & 0.159 \\
\hline
\end{tabular}
\end{table}

\begin{table}
\tiny
\caption{NGC 3962 surface photometry}
\label{table:n3962sb}
 \centering
  \begin{tabular}{@{}lcccc@{}}
    \hline\hline
$R$  &   $g$  & $g_{\rm err}$& $i$ & $i_{\rm err}$\\
\hline
5.82 & 18.990 & 0.000 & 17.844 & 0.001 \\
6.38 & 19.116 & 0.000 & 17.976 & 0.001 \\
6.98 & 19.239 & 0.000 & 18.104 & 0.001 \\
7.63 & 19.361 & 0.000 & 18.226 & 0.001 \\
8.34 & 19.478 & 0.001 & 18.344 & 0.001 \\
9.13 & 19.594 & 0.001 & 18.469 & 0.001 \\
10.10 & 19.725 & 0.001 & 18.595 & 0.001 \\
11.10 & 19.858 & 0.001 & 18.740 & 0.001 \\
12.29 & 20.010 & 0.001 & 18.900 & 0.002 \\
13.61 & 20.177 & 0.001 & 19.074 & 0.002 \\
15.07 & 20.355 & 0.001 & 19.257 & 0.002 \\
16.59 & 20.530 & 0.001 & 19.440 & 0.002 \\
18.26 & 20.710 & 0.002 & 19.630 & 0.003 \\
20.06 & 20.891 & 0.002 & 19.808 & 0.003 \\
21.97 & 21.061 & 0.002 & 19.989 & 0.004 \\
24.17 & 21.239 & 0.003 & 20.176 & 0.005 \\
26.66 & 21.430 & 0.003 & 20.366 & 0.006 \\
29.33 & 21.612 & 0.004 & 20.553 & 0.007 \\
32.23 & 21.787 & 0.004 & 20.725 & 0.008 \\
35.10 & 21.936 & 0.005 & 20.869 & 0.009 \\
38.18 & 22.069 & 0.006 & 21.000 & 0.010 \\
41.88 & 22.218 & 0.006 & 21.150 & 0.012 \\
46.26 & 22.375 & 0.007 & 21.319 & 0.014 \\
51.01 & 22.547 & 0.009 & 21.499 & 0.016 \\
55.94 & 22.723 & 0.010 & 21.678 & 0.019 \\
60.98 & 22.882 & 0.012 & 21.841 & 0.022 \\
66.48 & 23.067 & 0.014 & 22.031 & 0.027 \\
73.13 & 23.282 & 0.017 & 22.257 & 0.033 \\
80.44 & 23.526 & 0.021 & 22.489 & 0.040 \\
88.49 & 23.775 & 0.027 & 22.744 & 0.051 \\
97.33 & 24.028 & 0.033 & 22.921 & 0.059 \\
107.07 & 24.378 & 0.046 & 23.360 & 0.088 \\
117.78 & 24.712 & 0.062 & 23.611 & 0.109 \\
129.55 & 25.094 & 0.086 & 23.771 & 0.125 \\
142.51 & 25.323 & 0.107 & 24.316 & 0.201 \\
156.76 & 26.018 & 0.196 & 25.150 & 0.396 \\
172.43 & 26.870 & 0.395 & 26.267 & 0.875 \\
\hline
\end{tabular}
\end{table}

\begin{table}
\tiny
\caption{NGC 4240 surface photometry}
\label{table:n4240sb}
 \centering
  \begin{tabular}{@{}lcccc@{}}
    \hline\hline
$R$  &   $g$  & $g_{\rm err}$& $i$ & $i_{\rm err}$\\
\hline
1.23 & 17.672 & 0.000 & 17.039 & 0.000 \\
1.34 & 17.780 & 0.000 & 17.042 & 0.000 \\
1.47 & 17.915 & 0.000 & 17.045 & 0.000 \\
1.61 & 18.051 & 0.000 & 17.075 & 0.000 \\
1.76 & 18.183 & 0.000 & 17.180 & 0.000 \\
1.92 & 18.313 & 0.000 & 17.305 & 0.000 \\
2.09 & 18.438 & 0.000 & 17.427 & 0.000 \\
2.26 & 18.554 & 0.000 & 17.542 & 0.000 \\
2.44 & 18.664 & 0.000 & 17.650 & 0.000 \\
2.63 & 18.766 & 0.000 & 17.753 & 0.000 \\
2.84 & 18.863 & 0.000 & 17.858 & 0.000 \\
3.08 & 18.971 & 0.000 & 17.972 & 0.000 \\
3.35 & 19.091 & 0.000 & 18.102 & 0.000 \\
3.71 & 19.235 & 0.000 & 18.256 & 0.000 \\
4.21 & 19.417 & 0.000 & 18.449 & 0.000 \\
4.90 & 19.631 & 0.000 & 18.656 & 0.000 \\
5.54 & 19.835 & 0.000 & 18.852 & 0.000 \\
6.15 & 20.006 & 0.000 & 19.031 & 0.000 \\
6.79 & 20.170 & 0.000 & 19.202 & 0.000 \\
7.53 & 20.343 & 0.000 & 19.380 & 0.000 \\
8.32 & 20.515 & 0.000 & 19.555 & 0.000 \\
9.13 & 20.689 & 0.000 & 19.732 & 0.000 \\
10.02 & 20.893 & 0.000 & 19.937 & 0.000 \\
11.03 & 21.117 & 0.000 & 20.167 & 0.000 \\
12.16 & 21.344 & 0.000 & 20.399 & 0.000 \\
13.32 & 21.546 & 0.000 & 20.603 & 0.000 \\
14.62 & 21.736 & 0.000 & 20.798 & 0.000 \\
16.04 & 21.917 & 0.000 & 20.985 & 0.001 \\
17.58 & 22.092 & 0.000 & 21.174 & 0.001 \\
19.19 & 22.243 & 0.001 & 21.311 & 0.001 \\
20.99 & 22.415 & 0.001 & 21.523 & 0.001 \\
23.13 & 22.602 & 0.001 & 21.688 & 0.001 \\
25.34 & 22.780 & 0.001 & 21.880 & 0.001 \\
27.97 & 22.996 & 0.001 & 22.091 & 0.001 \\
30.94 & 23.256 & 0.001 & 22.384 & 0.002 \\
33.94 & 23.521 & 0.002 & 22.650 & 0.002 \\
37.39 & 23.813 & 0.002 & 22.960 & 0.003 \\
41.13 & 24.108 & 0.003 & 23.220 & 0.004 \\
45.50 & 24.411 & 0.004 & 23.618 & 0.006 \\
50.50 & 24.732 & 0.005 & 23.930 & 0.008 \\
55.89 & 25.041 & 0.007 & 24.277 & 0.011 \\
61.95 & 25.388 & 0.009 & 24.630 & 0.015 \\
69.30 & 25.777 & 0.013 & 24.969 & 0.021 \\
74.52 & 26.024 & 0.017 & 25.146 & 0.025 \\
\hline
\end{tabular}
\end{table}

\begin{table}
\tiny
\caption{IC 4889 surface photometry}
\label{table:ic4889sb}
 \centering
  \begin{tabular}{@{}lcccc@{}}
    \hline\hline
$R$  &   $g$  & $g_{\rm err}$& $i$ & $i_{\rm err}$\\
\hline

2.57 & 17.980 & 0.000 & 17.319 & 0.000 \\
2.83 & 18.092 & 0.000 & 17.319 & 0.000 \\
3.12 & 18.203 & 0.000 & 17.319 & 0.000 \\
3.41 & 18.324 & 0.000 & 17.320 & 0.000 \\
3.74 & 18.453 & 0.000 & 17.372 & 0.000 \\
4.11 & 18.581 & 0.000 & 17.502 & 0.000 \\
4.52 & 18.716 & 0.000 & 17.635 & 0.000 \\
4.98 & 18.849 & 0.000 & 17.768 & 0.000 \\
5.46 & 18.980 & 0.000 & 17.904 & 0.001 \\
6.00 & 19.123 & 0.000 & 18.042 & 0.001 \\
6.67 & 19.298 & 0.000 & 18.209 & 0.001 \\
7.28 & 19.428 & 0.000 & 18.338 & 0.001 \\
8.03 & 19.574 & 0.000 & 18.482 & 0.001 \\
8.75 & 19.699 & 0.000 & 18.677 & 0.001 \\
9.65 & 19.837 & 0.000 & 18.830 & 0.001 \\
10.64 & 20.009 & 0.000 & 18.930 & 0.001 \\
11.70 & 20.160 & 0.000 & 19.075 & 0.001 \\
12.79 & 20.303 & 0.001 & 19.216 & 0.002 \\
13.96 & 20.444 & 0.001 & 19.354 & 0.002 \\
15.27 & 20.594 & 0.001 & 19.502 & 0.002 \\
16.78 & 20.764 & 0.001 & 19.651 & 0.003 \\
18.32 & 20.927 & 0.001 & 19.812 & 0.003 \\
19.91 & 21.083 & 0.001 & 19.967 & 0.003 \\
21.75 & 21.247 & 0.001 & 20.135 & 0.004 \\
23.87 & 21.437 & 0.002 & 20.316 & 0.005 \\
26.52 & 21.664 & 0.002 & 20.508 & 0.006 \\
29.27 & 21.891 & 0.002 & 20.718 & 0.007 \\
32.38 & 22.132 & 0.003 & 20.932 & 0.008 \\
36.16 & 22.403 & 0.004 & 21.168 & 0.010 \\
40.36 & 22.687 & 0.005 & 21.411 & 0.013 \\
44.99 & 22.965 & 0.006 & 21.638 & 0.016 \\
50.01 & 23.222 & 0.008 & 21.845 & 0.019 \\
55.68 & 23.493 & 0.010 & 22.005 & 0.022 \\
61.74 & 23.748 & 0.013 & 22.235 & 0.027 \\
68.51 & 24.027 & 0.017 & 22.463 & 0.033 \\
79.26 & 24.486 & 0.025 & 22.616 & 0.038 \\
88.12 & 24.755 & 0.032 & 22.798 & 0.045 \\
97.77 & 25.153 & 0.046 & 23.011 & 0.054 \\
110.10 & 25.633 & 0.071 & 23.285 & 0.069 \\
122.16 & 26.178 & 0.115 & 23.613 & 0.093 \\
139.19 & 27.188 & 0.269 & 23.959 & 0.126 \\
\hline
\end{tabular}
\end{table}

\end{appendix}

\end{document}